\documentclass[preprint2]{proto}
\usepackage{times}
\usepackage{graphicx}
\usepackage{pdflscape}
\usepackage{color}

\begin{document}

\title{\textbf{\LARGE The Active Asteroids\footnote
{2015-February 8: This chapter is to appear in the book ASTEROIDS IV, part of the University of Arizona Space Science Series, edited by P. Michel, F. DeMeo and W. Bottke. 
}   
}}

\author {\textbf{\large David Jewitt}}
\affil{\small\em University of California at Los Angeles, USA}

\author {\textbf{\large Henry Hsieh}}
\affil{\small\em Academia Sinica, Taiwan}

\author {\textbf{\large Jessica Agarwal}}
\affil{\small\em Max Planck Institute for Solar System Research, Germany}

\begin{abstract}
\begin{list}{ } {\rightmargin 1in}
{\leftmargin -1in}
\baselineskip = 11pt
\parindent=1pc
{\small Some  asteroids  eject dust, producing transient, comet-like comae and tails; these are the active asteroids.  The causes of activity in this newly-identified population are many and varied. They include impact ejection and disruption, rotational instabilities, electrostatic repulsion, radiation pressure sweeping, dehydration stresses and thermal fracture, in addition to the sublimation of asteroidal ice.  These processes were either unsuspected or thought to lie beyond the realm of observation before the discovery of asteroid activity.    Scientific interest in the active asteroids lies in their promise to open new avenues into the direct study of  asteroid destruction, the production of interplanetary  debris, the abundance of asteroid ice and the origin of terrestrial planet volatiles.   
 \\~\\~\\~}
 \end{list}
\end{abstract}

\section{\textbf{INTRODUCTION}}

Small solar system bodies are conventionally labeled as either asteroids or comets, based on three distinct properties;   (1) Observationally, small bodies with unbound atmospheres (``comae'') are known as comets, while objects lacking such atmospheres are called asteroids.  (2) Dynamically, comets and asteroids are broadly distinguished by the use of a dynamical parameter, most commonly the Tisserand parameter measured with respect to Jupiter (\textit{Kresak} 1982, \textit{Kosai} 1992).   It is defined by

\begin{equation}
T_J = \frac{a_J}{a} + 2\left[(1-e^2)\frac{a}{a_J}\right]^{1/2}\cos(i)
\label{tisserand}
\end{equation}

\noindent where $a$, $e$ and $i$ are the semimajor axis, eccentricity and inclination of the orbit (relative to Jupiter's orbit) while $a_J$ = 5.2 AU is the semimajor axis of the orbit of Jupiter.  This parameter, which is conserved in the circular, restricted 3-body problem, provides a measure of the close-approach speed to Jupiter. Jupiter itself has $T_J$ = 3. Main belt asteroids have $a < a_J$ and  $T_J >$ 3 while dynamical comets (from the Kuiper belt and Oort cloud) have $T_J < 3$.   (3) Compositionally,  comets are ice-rich small bodies  formed beyond the snow-line in the protoplanetary disk, while the asteroids are ice-free and formed inside it.  

While there is often a reassuring concordance among these classification systems, all three are potentially  fallible.  For instance, the ability to detect a low surface brightness coma or tail is, in part, a function of  instrumental parameters and observing conditions.  The utility of Equation (\ref{tisserand})  is limited for objects with $T_J$ very close to 3 because the underlying criterion is based on an idealized representation of the Solar system (e.g.~Jupiter's orbit is not a circle, the gravity of other planets is not negligible, and non-gravitational forces due to outgassing and photon momentum can be important).   The least useful metric is the composition because, except in special cases, we have no practical way to measure either the composition of a small body and neither can we determine its formation location.

 \begin{figure}[h]
 \epsscale{01.0}
\plotone{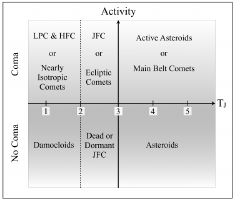}
 \caption{\small Empirical classification of small bodies based on the Tisserand parameter, $T_J$, (x-axis) and the presence or absence of coma (y-axis).  JFC, LPC and HFC are the Jupiter family, Long period and Halley family comet sub-types, distinguished by their dynamics. From Jewitt (2012). \label{classification}}  
 \end{figure}

Taken together, the observational and dynamical classifications suggest a simple two-parameter schematic that usefully describes the solar system's small body populations (Figure \ref{classification}).   The four quadrants in Figure (\ref{classification})  conveniently separate comets (upper left) from asteroids (lower right) and distinguish likely defunct comets (lower left) in which there is no activity presumably due to the past depletion of near-surface volatiles (Hartmann et al.~1987), from the newly recognized active asteroid class (upper right). 

\begin{figure}[h]
 \epsscale{1.0}
\plotone{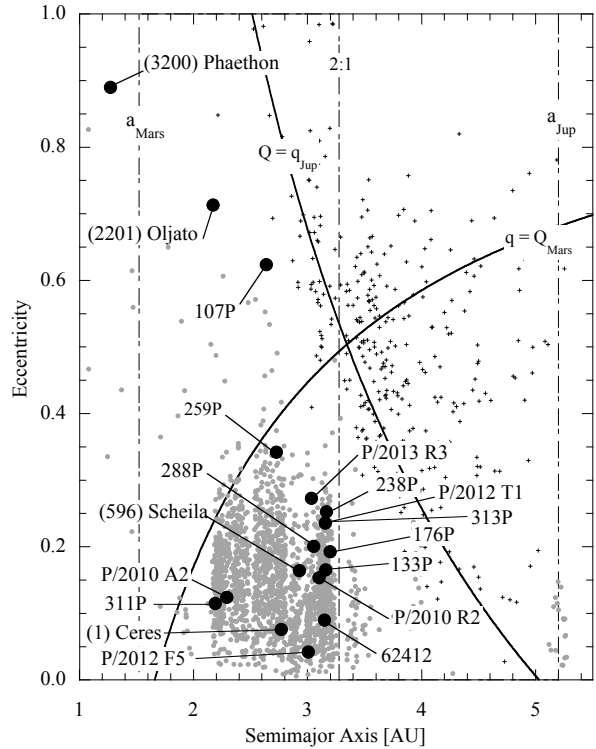}
 \caption{ Distribution of the  active asteroids in the semimajor axis vs.~orbital eccentricity plane.  Dynamical asteroids are shown as small grey circles, comets as small ``+'' symbols, and the active asteroids as black circles, each labeled with the object name (c.f.~Table \ref{orbital}).  Objects plotted above the diagonal arcs cross either the orbit of Mars, or of Jupiter. \label{ae}}  
 \end{figure}

The working definition employed here is that active asteroids are small bodies which 1) have semimajor axis $a < a_J$, 2)  have  $T_J >$ 3.08 and 3) show evidence for mass loss, for example in the form of a resolved coma or tail. In defining the critical Tisserand parameter separating asteroids from comets as $T_J$ = 3.08, rather than 3.0, we avoid many ambiguous cases caused by deviations of the real solar system from the circular, restricted 3-body approximation.   This definition also excludes Encke-type comets (2P/Encke has $T_J$ = 3.02), and the quasi-Hilda comets ($T_J \sim$ 2.9 - 3.04).  The orbital distribution of the currently known active asteroids is shown in Figure (\ref{ae}).  Three objects occupy planet-crossing orbits while the remaining 14 orbit in the main-belt.

The active asteroids are remarkable for being an entirely new population located in one of the closest and most intensively-studied regions of the solar system. Their activity is driven by a surprisingly diverse set of physical processes.  Reported examples include hypervelocity impact (e.g.~(596) Scheila), rotational instability probably driven by YORP torques (e.g.~311P, P/2013 R3), thermal disintegration driven by intense solar heating of minerals ((3200) Phaethon) and the sublimation of ice ((1) Ceres, 133P, 238P and 313P are the strongest examples).   Impact and rotational disruptions are primary sources of meteorites and larger planet impactors.  Observations promise to improve our understanding of the physics, and of the rates, of both asteroid destruction processes and meteoroid production. Analogous disruptions occurring around other stars are likely responsible for the creation of debris disks (e.g. \textit{Shannon and Wu} 2011).  A common dynamical end-fate of planet-crossing bodies is to strike the Sun; a better understanding of the role of thermal disintegration will be important both in the solar system and in the context of photospheric impactors in polluted white dwarf systems (\textit{Jura and Xu} 2013).  Lastly, the survival of primordial ice in the asteroids may offer the opportunity to sample volatiles from a region of the protoplanetary disk different from that in which the Kuiper belt and Oort cloud comets formed.  The outer asteroid belt is also a likely source region for the volatile inventory of the Earth (\textit{Morbidelli et al.,}~2000; \textit{O'Brien et al.,} 2006), giving new relevance to the origin of the oceans. 

The active asteroids were called ``main-belt comets'' by \textit{Hsieh and Jewitt} (2006) while \textit{Hsieh et al.,} (2012a) employed the term ``disrupted asteroids'' to refer to objects that exhibit comet-like activity believed to be non-sublimation-driven.  We use the more general term ``active asteroids'' both because some of the known examples  are not in the main-belt (c.f.~Figure \ref{ae}), and because this nomenclature implies no supposition about the cause of the activity.

The active asteroids were reviewed by \textit{Bertini} (2011) and, in more detail, by \textit{Jewitt} (2012).  This chapter updates the latter paper with many new observations and examples of asteroid activity and adopts a tutorial style in the interests of clarity.

\section{\textbf{CURRENTLY KNOWN ACTIVE ASTEROIDS}}

We list the active asteroids in the order of decreasing Tisserand parameter in Table (\ref{orbital}) and discuss them briefly here.

\subsection{(3200) Phaethon, $T_J$ = 4.508}

\begin{figure}[h]
 \epsscale{1.0}
 \plotone{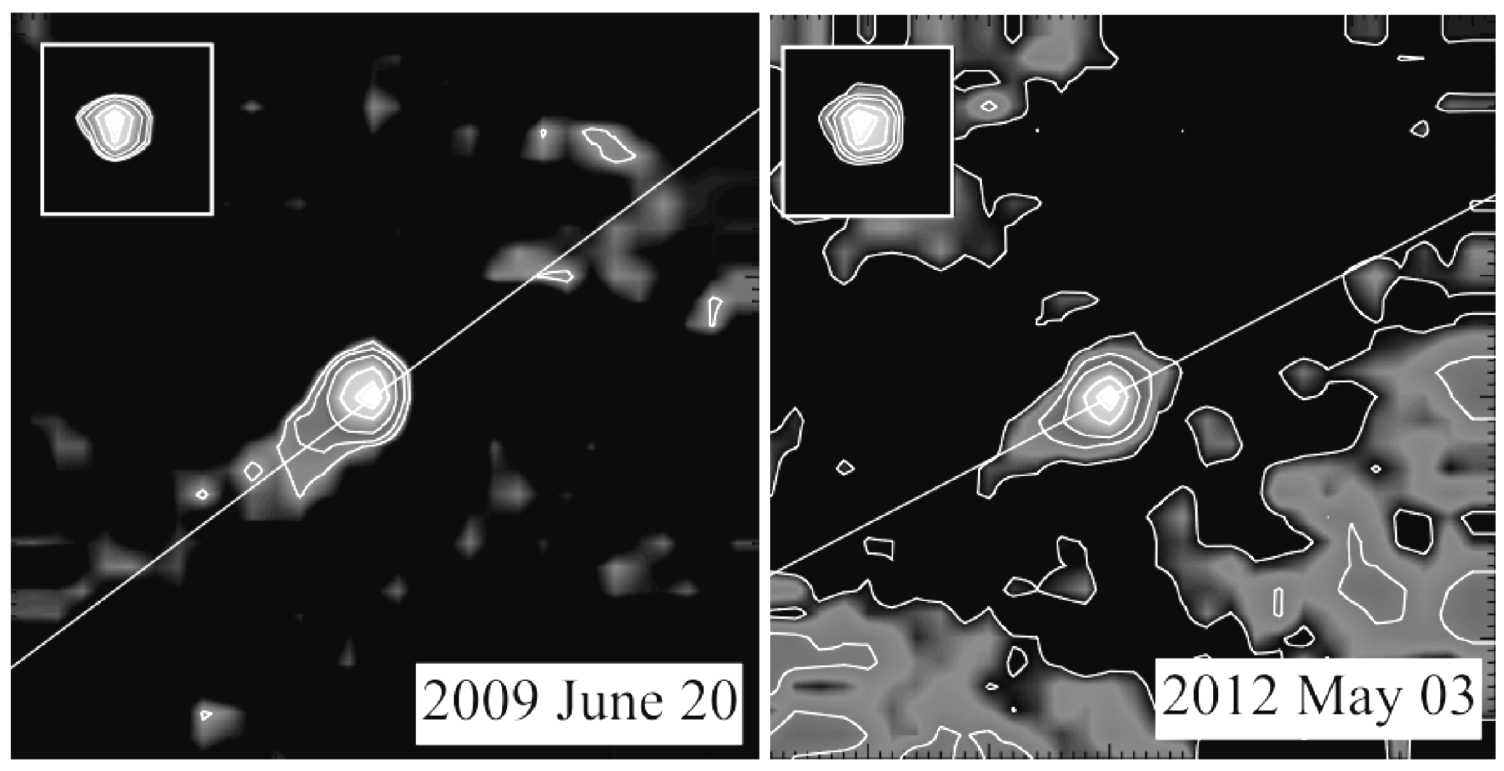}
 \caption{\small (3200) Phaethon at perihelion in 2009 and 2012 showing extended emission along the projected Sun-comet line.  The insets show the point-spread function of the STEREO camera.  Each panel shows a region 490\arcsec square and is the median of $\sim$30 images taken over a 1 day period.  From Jewitt et al.~2013.\label{phaethon}}  
 \end{figure}

Phaethon is a B-type asteroid with a 0.14 AU perihelion distance, a  dynamical lifetime to scattering by the terrestrial planets $\sim$100 Myr and a source in the main asteroid belt that may be related to (2) Pallas (\textit{de Leon et al.,} 2010).   This 5 km diameter body is the source of the Geminid meteor stream (e.g. \textit{Williams and Wu} 1993) and is dynamically related to kilometer-sized asteroids 2005 UD (\textit{Ohtsuka et al.}~2006, \textit{Jewitt and Hsieh} 2006, \textit{Kinoshita et al.}~2007) and 1999 YC (\textit{Kasuga and Jewitt} 2008).   The dispersion age of the Geminid stream is $t \sim$10$^3$ yrs (\textit{Ohtsuka et al.}~2006), meaning that Phaethon is active on this or a shorter timescale.  The timescale for the separation of 2005 UD and 1999 YC is not known but, presumably, is much longer.    No gas has been reported in optical spectra (\textit{Chamberlin et al.,} 1996) but near-Sun brightening of Phaethon by a factor of two was detected in STEREO spacecraft data  in 2009 (\textit{Jewitt and Li} 2010) and 2012 (\textit{Li and Jewitt} 2013), while the ejected dust has also been resolved (\textit{Jewitt et al.,} 2013, c.f.~Figure \ref{phaethon}).  However, the sudden appearance and position angle of the Phaethon dust tail indicate that the ejected particles are small, with an effective radius $\sim$1 $\mu$m and a combined mass $\sim$3$\times$10$^5$ kg (\textit{Jewitt et al.,} 2013).  This is tiny compared to the $M_s \sim$ 10$^{12}$ to 10$^{13}$ kg Geminid stream mass (\textit{Hughes and McBride} 1989; \textit{Jenniskens} 1994) and suggests that the Geminids are produced by a different process. In any case, 1 $\mu$m particles are quickly accelerated by solar radiation pressure to faster than the solar system escape speed, and cannot contribute to the Geminid stream.  Larger particles evidently contribute too little to the optical scattering cross-section to be discerned in STEREO near-Sun observations taken against the bright coronal background.  However, particles with sizes $>$10 $\mu$m were recently reported in thermal emission at 25 $\mu$m (\textit{Arendt} 2014), while kilogram-mass Geminids have been recorded striking the night-side of the Moon (\textit{Yanagisawa et al.,} 2008).  Some such bodies might survive passage through the Earth's atmosphere  (\textit{Madiedo et al.,} 2013) and could already be present, but unrecognised, in terrestrial meteorite collections.

The Geminid stream mass and dynamical age together could imply ejection of debris from Phaethon, if in steady-state, at rates 30 $\lesssim M_s/t \lesssim$ 300 kg s$^{-1}$.  More likely, mass loss  from Phaethon is highly variable, with dramatic bursts  interspersed with long periods of quiescence.  Continued observations of Phaethon, especially at long wavelengths sensitive to large particles, are needed.  

\subsection{311P/PANSTARRS (P/2013 P5), $T_J$ =  3.662}

\begin{figure}[h]
 \epsscale{0.99}
\plotone{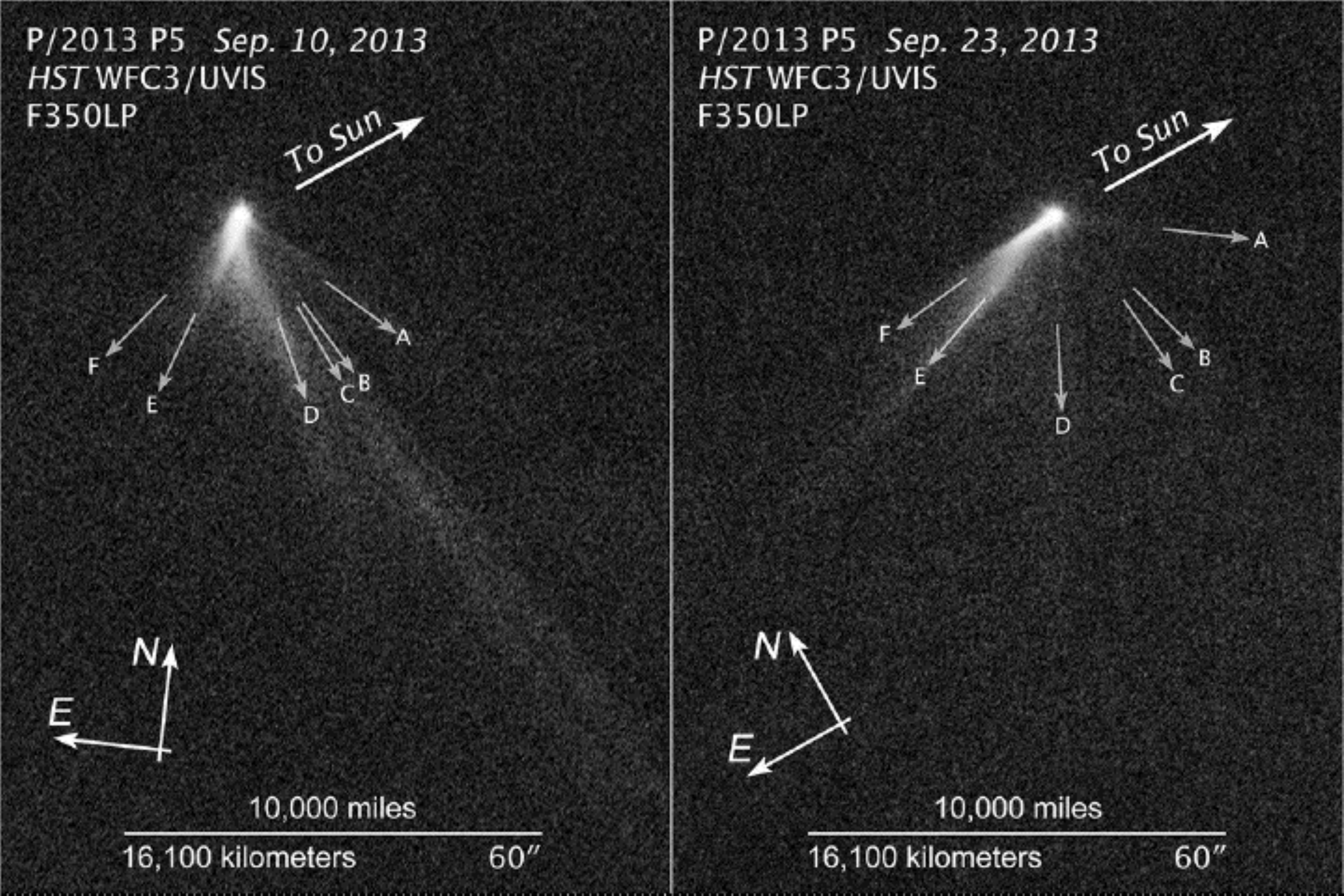}
  \caption{\small 311P observed on two epochs showing the distinctive evolving, multiple tail structure.  From \textit{Jewitt et al.,} (2013). \label{311P}}  
 \end{figure}

311P is an inner belt asteroid (Table \ref{orbital}) that ejected dust episodically over at least nine months in 2013, creating a remarkable multi-tail appearance (\textit{Jewitt et al.,} 2013, 2014; \textit{Hainaut et al.,} 2014; \textit{Moreno et al.,} 2014, c.f.~Figure \ref{311P}).  Interpreted as synchrones (the sky-plane projected positions of dust particles of different sizes released simultaneously from the nucleus), each  tail has a position angle linked to the ejection date.  The intervals between ejections appear random.

The episodic mass loss is unlike that seen in any previously observed comet.  This fact alone argues against ice sublimation as the driving agent.  An additional consideration is that the orbit of 311P lies near the inner edge of the asteroid belt, in the vicinity of the Flora family. The Floras have been associated with the LL chondrites (\textit{Vernazza et al.,} 2008), which themselves reflect metamorphism to temperatures $\sim$800\degr C to 960\degr C (\textit{Keil} 2000).  It is improbable that water ice could  survive in such a body.   Impact likewise offers an untenable explanation for activity that occurs episodically over many months.

The color of 311P indicates an S-type classification (\textit{Jewitt et al.,} 2013, \textit{Hainaut et al.,} 2014), consistent with its inner-belt orbit and with the Floras.   Flora family asteroids have a mean visual geometric albedo 0.29$\pm$0.09 (\textit{Masiero et al.,} 2013).  With this assumed albedo, the nucleus of 311P has a radius $r_n \le$ 240$\pm$40 m (\textit{Jewitt et al.,} 2013).   This small size, combined with the inner-belt location, renders 311P susceptible to spin-up by radiation forces.  Specifically, the YORP timescale for 311P is $<$10$^6$ yr, shorter than the collisional lifetime.  Therefore, it is reasonable to conjecture that episodic mass loss from 311P results from a rotational instability in which regolith is locally unstable and occasionally avalanches off the surface in response to rapid spin (\textit{Jewitt et al.,} 2013) .  This ``rotational mass shedding'' qualitatively accounts for the non-steady tail formation and the success of synchrone models, which assume ejection from the nucleus  at zero initial velocity.  Unstable material on 311P would depart at the gravitational escape speed, $V_e \sim$ 0.3 m s$^{-1}$ for a body with $\rho$ = 3300 kg m$^{-3}$, the density of the LL chondrites, or less.   \textit{Hainaut et al.} (2014) suggest an alternative model in which dust production results from friction between two oscillating components of a contact binary nucleus.

\subsection{P/2010 A2 (LINEAR), $T_J$ =  3.582}

\begin{figure}[h]
 \epsscale{0.99}
\plotone{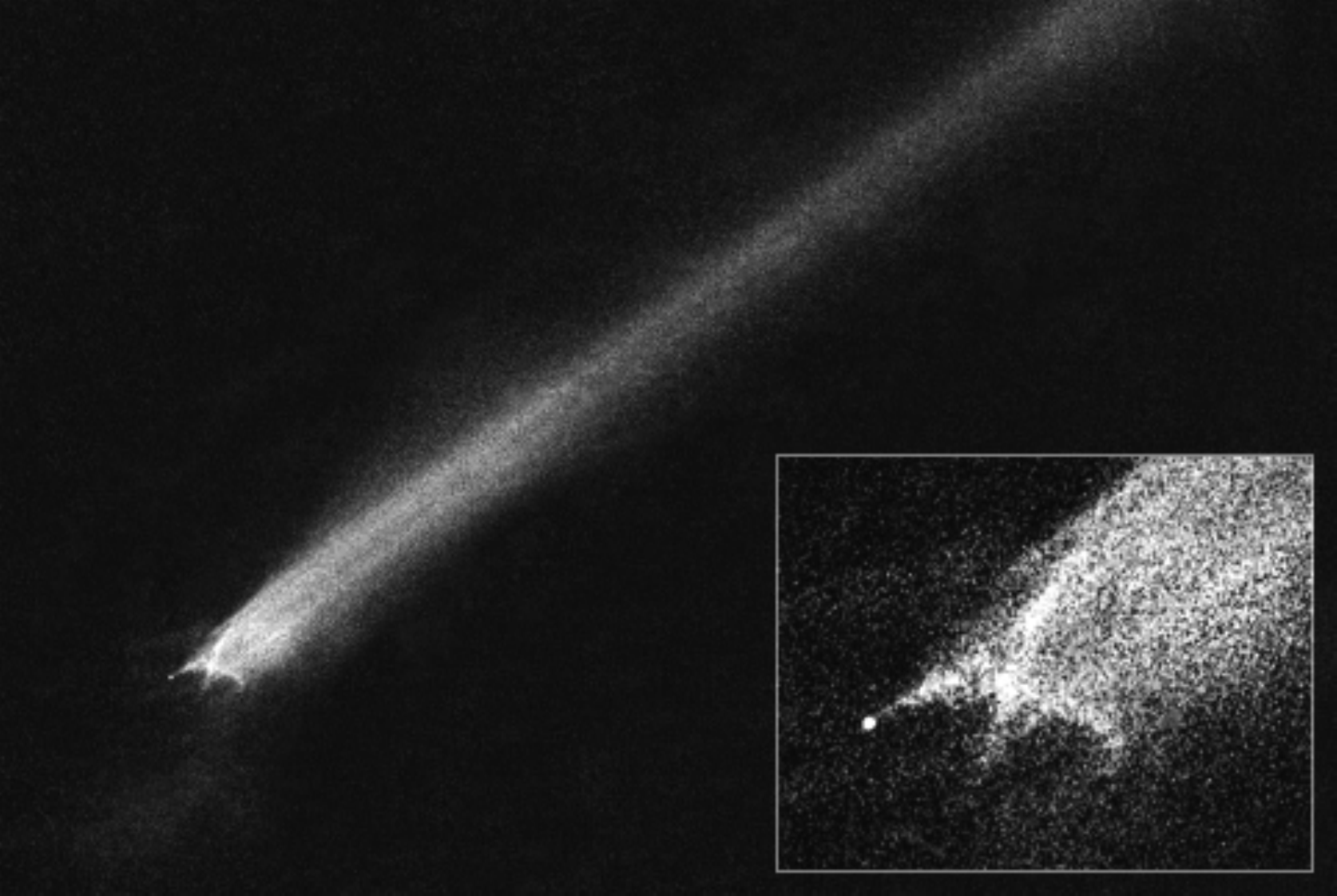}
  \caption{\small Trailing structures in P/2010 A2 observed UT 2010 January 29.  The full image width is $\sim$1\arcmin~while the inset width is $\sim$12\arcsec.  From \textit{Jewitt et al.,} (2010).\label{2010A2}}  
 \end{figure}

The object showed a distinctive morphology with a leading, point-like nucleus about 120 m in diameter (Table \ref{physical}), trailed by an extended tail of dust in which are embedded ribbon-like structures (Jewitt et al.~2010, see Figure \ref{2010A2}).  The position angle of the tail and its variation with time are consistent with the action of radiation pressure on mm to cm sized dust particles, following impulsive ejection at very low speeds ($\sim$0.2 m s$^{-1}$) in 2009 February - March, nearly a year before discovery (\textit{Jewitt et al.,}~2010, \textit{Snodgrass et al.,}~2010).    Prediscovery observations were found  as early as UT 2009 November 22 while detection in the first $\sim$6 months after the dust ejection event  was impeded by the  angular proximity of P/2010 A2 to the Sun (\textit{Jewitt et al.,}~2011a).  Before discovery, a large quantity of fast-moving particles are presumed to have left the vicinity of the main nucleus.  The mass of particles remaining in the tail at discovery is estimated to be in the range (6 to 60)$\times$10$^7$ kg (\textit{Jewitt et al.,}~2010, \textit{Moreno et al.,}~2010, \textit{Snodgrass et al.,}~2010).  Observations in 2012 October reveal a surviving  trail of particles up to 20 cm in radius and a differential power law size distribution index 3.5$\pm$0.1 (\textit{Jewitt et al.,} 2013b).  \textit{Kim et al.,} (2012) found that the colors most closely resemble those of an H5 chondrite. The estimated total mass is $\sim$5$\times$10$^8$ kg, about 10\% of the mass of the nucleus.

The properties of P/2010 A2 appear consistent both with an impact origin (with a meter-scale projectile striking at 5 km s$^{-1}$) and with rotational instability (\textit{Jewitt et al.,} 2010), although an impact origin is often assumed (\textit{Snodgrass et al.,} 2010, Hainaut et al., 2012).  In the former interpretation, the distinctive arms seen in Figure (\ref{2010A2}) are projections of the impact cone (\textit{Kleyna et al.,} 2013) while, in the latter, they are ribbons of debris shed in the rotational equator (\textit{Agarwal et al.,} 2013).

\subsection{(1) Ceres, $T_J$ =  3.309}

Ceres is unique among the known active asteroids in that it is the only object for which water sublimation has been spectroscopically detected.
The 556.936 GHz water ground-state line  was reported in absorption by \textit{K{\"u}ppers et al.,} (2014).  The line area is time-variable, and can be fitted by a model-dependent water production rate $dM/dt \sim$ 6 kg s$^{-1}$.  Unlike  other active asteroids, however, no dust emission has ever been reported for Ceres.

At the subsolar point on Ceres (heliocentric distance 2.6 AU) a perfectly absorbing water ice surface would sublimate in equilibrium with sunlight at the specific rate $f_s$ = 5$\times$10$^{-5}$ kg m$^{-2}$ s$^{-1}$ (c.f.~Section \ref{ice}). An exposed, subsolar ice patch of area $(dM/dt)/f_s$ =  0.12 km$^2$, corresponding to a circle of radius 200 m, could supply the measured water vapor.   A larger area of exposed ice would be needed if the ice were more reflective, or if it were located away from the subsolar point.  One possibility is that ice has been recently exposed on Ceres by a surface disturbance, possibly by the formation of a small impact crater or other geological instability.  Alternatively, the  water vapor might not be produced by sublimation in sunlight, but by subsurface heating followed by escape through a vent (c.f.~Enceladus; \textit{Porco et al.,} 2014), since  Ceres is potentially large enough to maintain significant internal heat (\textit{Castillo-Rogez and McCord} 2010). Additional measurements, presumably from the DAWN mission, will be needed to distinguish between these possibilities.

\subsection{(2201) Oljato, $T_J$ =  3.299}

Magnetometers on the Pioneer Venus spacecraft revealed, in the 1980s, multiple, symmetric interplanetary magnetic field enhancements, clumped non-randomly in time (Russell et al.~1984).  About 25\% of these events are associated with planet-crossing asteroid (2201) Oljato, whose orbit lies interior to Venus' when near perihelion.  Russell et al.~suggested that the magnetic disturbances result from deceleration of the solar wind, perhaps caused by mass loading from ionized gases released by an unknown process from debris distributed along Oljato's orbit.  A mass loading rate of only $\sim$5 kg s$^{-1}$ is reportedly needed.   Observations in 2006-2012 with the Venus Express under a similar geometry reveal no events related with Oljato.  \textit{Lai et al.,} (2014) argue that the field enhancements were due to loading of the interplanetary wind by charged nano-scale dust in Oljato's orbit and that the quantity of this dust decreased between the Pioneer and Venus Express missions.  
A spectroscopic search for gas produced by Oljato itself proved negative (Chamberlin et al.~1996), with upper limits to the CN production rate near 10$^{23}$ s$^{-1}$.  With a standard H$_2$O/CN mixing ratio of 360 (\textit{A'Hearn et al.,} 1995), the corresponding limit to the mass production rate in water is $\le$1.5 kg s$^{-1}$.  Whatever the cause of the repetitive magnetic disturbances, they are not products of an inert asteroid and imply mass loss from Oljato. A dynamical simulation indicates that (2201) Oljato has negligible chance of being a captured Jupiter family comet (\textit{Bottke et al.~2002}).

\subsection{P/2012 F5 (Gibbs), $T_J$ =  3.228}

\begin{figure}[h]
 \epsscale{0.99}
\plotone{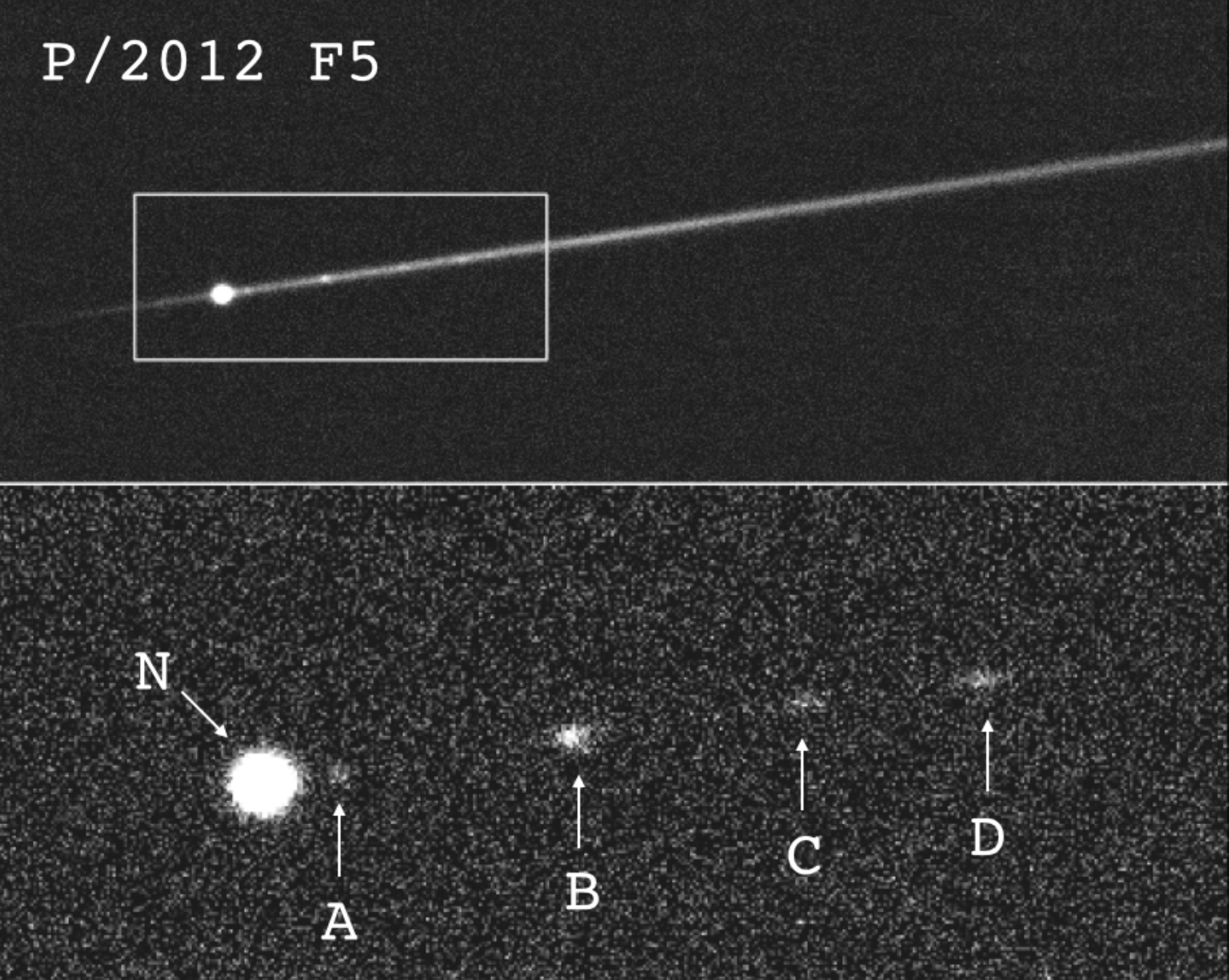}
  \caption{\small Keck telescope images of P/2012 F5 on UT 2014 August 26.   The region shown is 1.0$\times$2.5\arcmin. The rectangular box in the top panel shows the region enlarged in the bottom panel, where a model of the trail has also been subtracted. Letters mark  the primary nucleus, N, and tail condensations, A - D.  From \textit{Drahus et al.,} (2015).\label{2012F5}}  
 \end{figure}

P/2012 F5 was observed on 2012 September 18 to exhibit a dust trail extending $>15'$ in the plane of the sky while at a heliocentric distance of $R\sim3.1$~AU.  Follow-up observations in 2013 showed the object to be largely inactive and set an upper limit on the diameter of the nucleus of $\sim$2~km, although residual dust contamination of the nucleus photometry at the time could not be completely ruled out (\textit{Novakovi\'c et al.}, 2014).   A series of deep-images in 2014 showed rapid nucleus rotation (period 3.24 hr) and revealed four condensations in an orbit-aligned dust trail (Figure \ref{2012F5}).  Dynamical analysis of the object found it to be dynamically stable over at least 1~Gyr, and therefore unlikely to be recently implanted from elsewhere in the solar system (\textit{Stevenson et al.}, 2012), although it was also found to be a member of an extremely compact asteroid cluster determined to be just $1.5\pm0.1$~Myr in age (\textit{Novakovi\'c et al.}, 2014). 

The origin of the mass loss is unclear, with evidence consistent  both with impact and  rotational instability.  The dust trail was determined via numerical modeling to consist of particles ejected in a single impulsive event, consistent with impact, roughly nine months prior to the discovery of activity  (\textit{Moreno et al.}, 2012; \textit{Stevenson et al.}, 2012).   In contrast, the rapid rotation suggests that material may have been lost from F5 by rotational instability.  Future observations are needed to determine the ejection times and fates of the trail condensations.

\subsection{259P/(Garradd) (P/2008 R1), $T_J$ =  3.216}

259P was observed over a $\sim$45 day interval in 2008 to have the appearance of an active comet with a typical flared tail while at a heliocentric distance of $R\sim2$~AU (Jewitt et al., 2009).  The object's intrinsic brightness decreased by a factor of about 2 over the course of those observations, corresponding to a mass loss rate on the order of $\sim10^{-2}$~kg~s$^{-1}$, assuming mean grain radii of 10~$\mu$m and bulk densities of $\rho=1300$~kg~m$^{-3}$ (MacLennan \& Hsieh, 2012).  Subsequent observations of the inactive nucleus found an effective nucleus radius of $r=0.30\pm0.02$~km (Table \ref{physical}; MacLennan \& Hsieh 2012), assuming a red geometric albedo of 0.05.  Spectral observations limited the production of the CN radical to Q$_{CN} \le$ 1.4$\times$10$^{23}$ s$^{-1}$, corresponding to a water production rate $\le$1.5 kg s$^{-1}$ assuming H$_2$O/CN = 360.  259P is located near the 8:3 mean-motion resonance with Jupiter and is also affected by the $\nu_6$ secular resonance.   The dynamical lifetime in this orbit is short (20 to 30 Myr) compared to the age of the solar system, suggesting that 259P was scattered into its present location from elsewhere in the asteroid belt, or possibly even from elsewhere in the solar system.

\subsection{(596) Scheila, $T_J$ =  3.208}

\begin{figure}[h]
 \epsscale{0.90}
\plotone{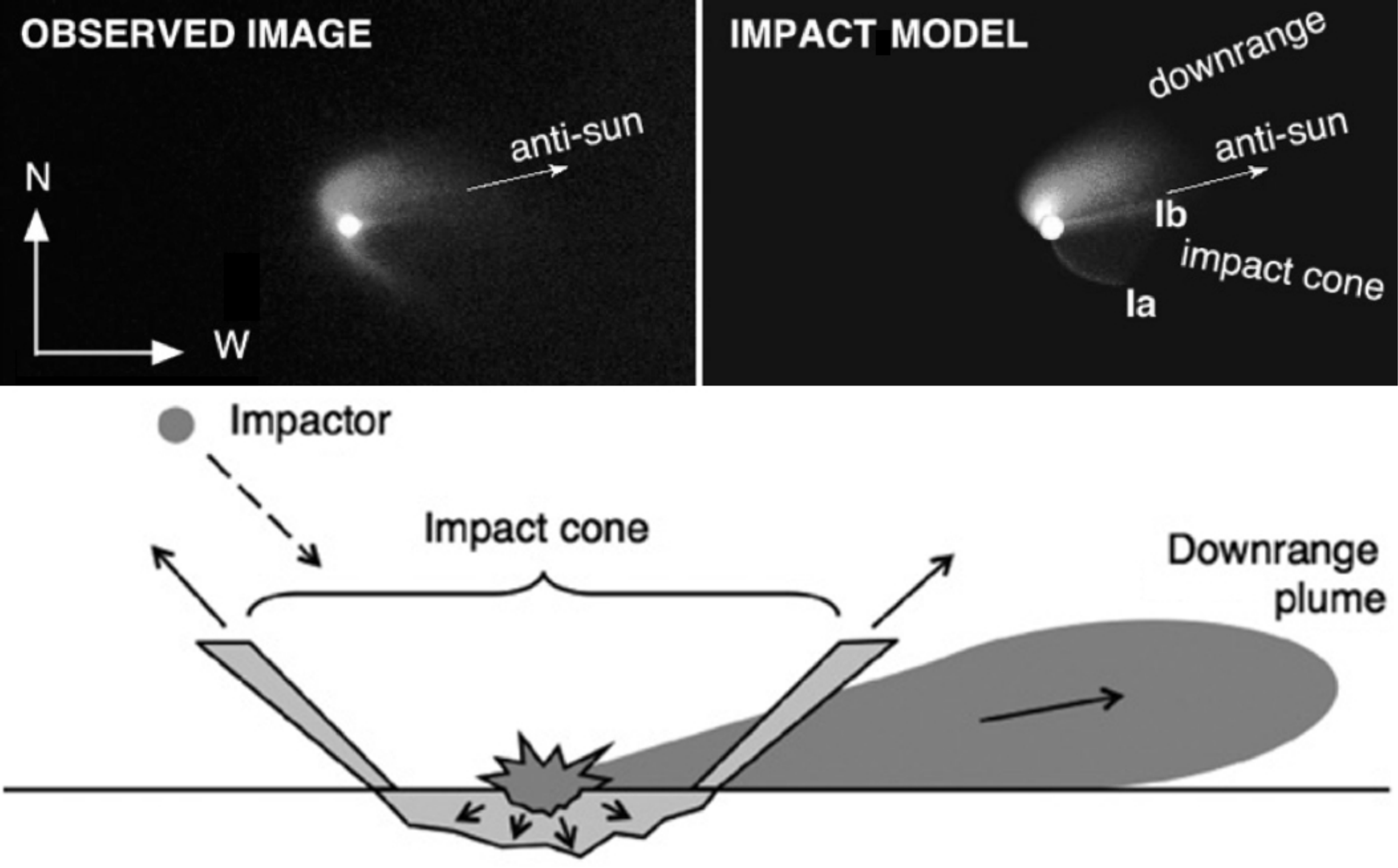}
  \caption{\small Observed image of Scheila on 2010 December 12 (top left), simulated image of dust ejecta consisting of an impact cone and downrange plume (top right), and diagram of the modeled impact scenario (bottom).  From \textit{Ishiguro et al.,} (2011b).\label{Scheila}}  
 \end{figure}

(596) Scheila, a 113 km diameter object with red geometric albedo $\sim$0.04 (Table \ref{physical}), developed a comet-like appearance in late 2010 in the form of two prominent dust plumes.  Over the course of a month, dust in these plumes dispersed from the nucleus due to solar radiation pressure, apparently without any continued replenishment of particles from the nucleus (\textit{Bodewits et al.}, 2011; \textit{Jewitt et al.}, 2011b; \textit{Moreno et al.}, 2011a).  Dust modeling by \textit{Ishiguro et al.}\ (2011a,b) demonstrated that the morphology of the observed dust plumes was consistent with the results of an impact-driven ejecta cloud consisting of a circularly-symmetric ejecta cone (subsequently inverted by radiation pressure) and a down-range plume (Figure~\ref{Scheila}). \textit{Bodewits et al.}, (2014) subsequently reported a change in the rotational lightcurve which they attributed to the signature of the impact scar.

Upper limits to the gas production from the nucleus of $Q_{OH}\le$10$^{26}$~s$^{-1}$ (corresponding to a water production rate of $<3$~kg~s$^{-1}$; \textit{Howell and Lovell}, 2011) and $Q_{\rm CN}<9\times10^{23}$ (corresponding to $Q_{\rm H_2O}\lesssim10$~kg~s$^{-1}$; \textit{Hsieh et al.}, 2012a) were found, though the meaning of these limits is unclear given the apparently impulsive nature of the mass loss event from Scheila.  \textit{Jewitt et al.}\ (2011) calculated the mass of dust in micron-sized grains to be 4$\times$10$^7$~kg, while more model-dependent attempts to account for larger particles gave total dust masses of $\sim10^8-10^{10}$~kg (\textit{Bodewits et al.}, 2011; \textit{Ishiguro et al.}, 2011a; \textit{Moreno et al.}, 2011).  No ice was observed in the coma (\textit{Yang and Hsieh}, 2011).

\subsection{288P/(300163) 2006 VW$_{139}$, $T_J$ =  3.203}

\begin{figure}[h]
 \epsscale{1.00}
\plotone{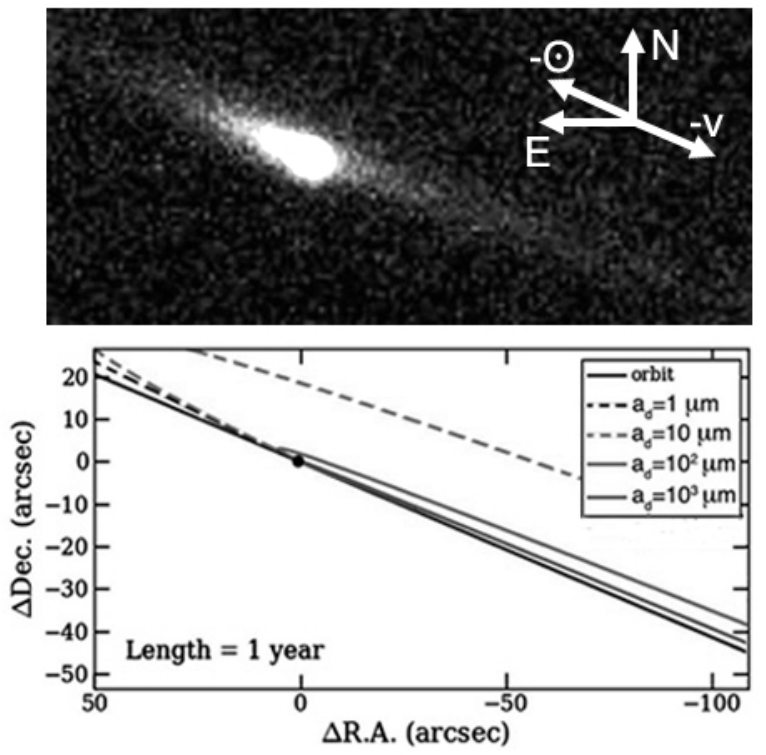}
  \caption{\small Image of 288P (top) and corresponding syndyne plot (bottom).  A syndyne is the locus of positions of particles of one size released from the nucleus at zero speed over a range of times.  From \textit{Hsieh et al.} (2012b).\label{288P}}
 \end{figure}

Discovered in 2006 as an inactive asteroid, 288P was found to be cometary in 2011 by the Pan-STARRS1 survey telescope (\textit{Hsieh et al.}, 2012b, c.f.~Figure \ref{288P}).  A short ($\sim10''$) antisolar dust tail and a longer ($\sim60''$) dust trail aligned with the object's orbit plane were seen in deep follow-up images, indicating the simultaneous presence of both recent and months-old dust emission, strongly suggesting that the observed activity was due to a long-duration emission event, consistent with sublimation.  Photometric monitoring showed that intrinsic brightness of the near-nucleus coma remained constant for at least one month, before then declining by 40\% over the next month, again consistent with a long-duration emission event where the coma was continually replenished by fast-dissipating small dust particles over its period of constant brightness, and then faded quickly once the replenishment rate slowed.

Spectroscopic observations did not detect any gaseous sublimation products, setting an upper limit to the CN production rate of $Q_{\rm CN}<10^{24}$~mol~s$^{-1}$ (\textit{Hsieh et al.}, 2012b; \textit{Licandro et al.}, 2013).  However, dust modeling found that the onset of activity occurred shortly after perihelion, and persisted for about 100 days (\textit{Licandro et al.}, 2013).  A dynamical analysis by \textit{Novakovi\'c et al.}\ (2012) found that 288P belongs to a compact cluster of 24 asteroids believed to originate from the fragmentation of a $\sim11$-km-diameter parent body $7.5\pm0.3$~Myr ago.

\subsection{(62412) 2000 SY178, $T_J$ =  3.197}

(62412) 2000 SY178 has diameter 7.8$\pm$0.6 km, geometric albedo 0.065$\pm$0.010 and is a probable C-type  (\textit{Sheppard and Trujillo} 2015). The nearly circular orbit at 3.146 AU is consistent with membership of the 2 to 3 Gyr old Hygiea family.    While the origin of the activity is unknown, the 3.33 hr rotation period and 0.45 magnitude lightcurve range suggest, by analogy with 133P (\textit{Jewitt et al.,} 2014a), that rotation may play a role.

\subsection{P/2013 R3 (Catalina-PANSTARRS), $T_J$ =  3.185}

\begin{figure}[h]
 \epsscale{0.90}
\plotone{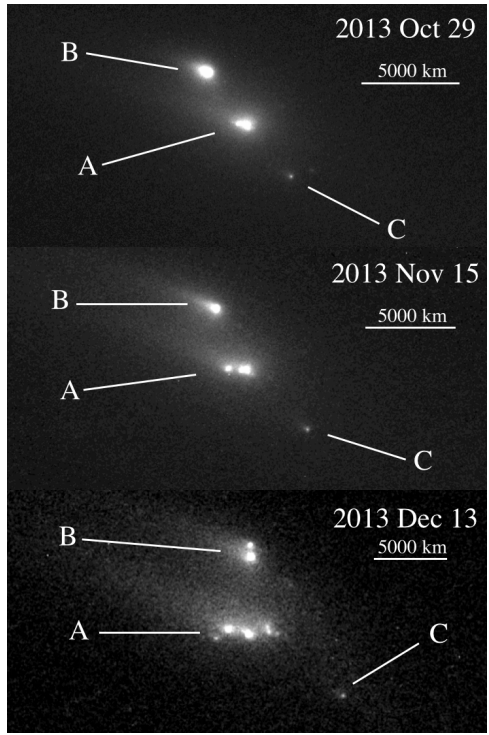}
  \caption{\small Evolution of the  components of P/2013 R3 over six weeks in 2014.  The components, labeled A, B and C, show progressive splitting and are enveloped in a 10$^8$ kg debris cloud.  Scale bars mark 5000 km at the distance of the object.  Adapted from \textit{Jewitt et al.,} (2014b).\label{2013R3}}  
 \end{figure}

P/2013 R3 is a dust enshrouded outer-belt asteroid observed in a state of disintegration (\textit{Jewitt et al.,} 2014b, c.f.~Figure \ref{2013R3}).  Ten distinct components were detected in the interval from 2014 October to December, with a velocity dispersion between fragments of order 0.3 to 0.5 m s$^{-1}$.  Because of dust contamination in 2014 it is only possible to set upper limits to the size of the fragments.  Assuming geometric albedo 0.05, the four largest have radii $\le$ 200 m.  The gravitational escape velocities of the largest fragments (assuming density $\rho$ = 1000 to 3000 kg m$^{-3}$) are 0.15 to 0.25 m s$^{-1}$, comparable to the measured fragment velocity dispersion.  The fragmentation took place successively over a period of several months at least, excluding an impact as the cause. Disruption due to the pressure of a sub-surface volatile reservoir was also excluded, although the continued dust activity of the fragments may have been caused by the sublimation of newly exposed ice. Rotational breakup seems the most likely cause of the catastrophic disruption of R3, supported by the close-to-escape speed relative velocities of the fragments (\textit{Hirabayashi et al., 2014b}).

\subsection{133P/(7968) Elst-Pizarro, $T_J$ =  3.184}

133P was first observed to be active in 1996, exhibiting a long, narrow dust trail with no visible coma, making it the first known (and currently best-characterized) active asteroid in the main asteroid belt.  At first suspected to be the product of a collision (\textit{Boehnhardt et al.,} 1998, \textit{Toth} 2000), it has since been observed to be active on three additional occasions in 2002, 2007, and 2013, and  inactive inbetween) (e.g., \textit{Hsieh et al.}, 2004, 2010; \textit{Jewitt et al.}, 2014a).

 \begin{figure}[h]
 \epsscale{0.90}
\plotone{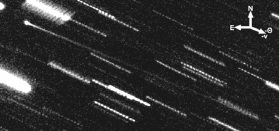}
  \caption{\small 133P imaged UT 2002 September 07.  From \textit{Hsieh et al.,} (2004).\label{133P}}  
 \end{figure}

The appearance of 133P when active is typically that of a point-like nucleus with a thin tail (or ``trail'') of  dust following in the projected orbit (Figure \ref{133P}).  Order of magnitude dust mass loss rates, inferred from surface photometry of the tail (\textit{Hsieh et al.}, 2004) reach $\sim$0.02~kg~s$^{-1}$, while comparable upper limits in gas are inferred from spectroscopy (\textit{Licandro et al.}, 2011a).  The thin tail indicates that particles are ejected very slowly from the nucleus with characteristic speeds $\sim$1.8 $a_{\mu m}^{-1/2}$~m~s$^{-1}$, where $a_{\mu m}$ is the grain radius expressed in microns (\textit{Jewitt et al.,} 2014a).  Curiously, particles larger than a few microns leave the nucleus at speeds below the $\sim$2 m s$^{-1}$ gravitational escape speed (e.g.~mm sized particles have $v \sim$ 6 cm s$^{-1}$).  This is inconsistent with a pure sublimation origin, and appears to require centripetal assistance and a spatially subdivided sublimating surface (\textit{Jewitt et al.,} 2014a).  

The nucleus has a rotation period of $3.471\pm0.001$~hr (\textit{Hsieh et al.}, 2004), a spectrum similar to those of B- or F-type asteroids (e.g., \textit{Bagnulo et al.}, 2010; \textit{Licandro et al.}, 2011a), and an optical albedo of $p_R\sim p_V\sim0.05$ (\textit{Hsieh et al.}, 2009; \textit{Bauer et al.}, 2012).

\subsection{176P/(118401) LINEAR (118401), $T_J$ =  3.167}

176P/LINEAR was observed to exhibit a short fan-shaped tail over a single, month-long interval in 2005 (\textit{Hsieh et al.}, 2011a).  During this time, the object was about 30\% brighter than the bare nucleus, leading to an implied dust mass $\sim$10$^5$ kg.  The properties of the dust can be approximately matched by models in which the characteristic particle size is 10 $\mu$m, the ejection speed $\sim$5 m s$^{-1}$ and the dust production rate $\sim$0.07 kg s$^{-1}$, all similar to values inferred in 133P.  Activity was sought but not detected in 2011 (\textit{Hsieh et al.,} 2014) and thee water production rate was spectroscopically limited to $Q_{H_2O} \le$ 4$\times$10$^{25}$ s$^{-1}$ ($\le$ 1.0 kg s$^{-1}$) in Herschel observations 40 days after perihelion on UT 2011 August 8 (\textit{de Val-Borro} 2012). The 4.0$\pm$0.4 km diameter nucleus rotates with a period near 22.2 hr (Table \ref{physical}).

\subsection{238P/Read, $T_J$ =  3.152}

\begin{figure}[h]
 \epsscale{0.90}
\plotone{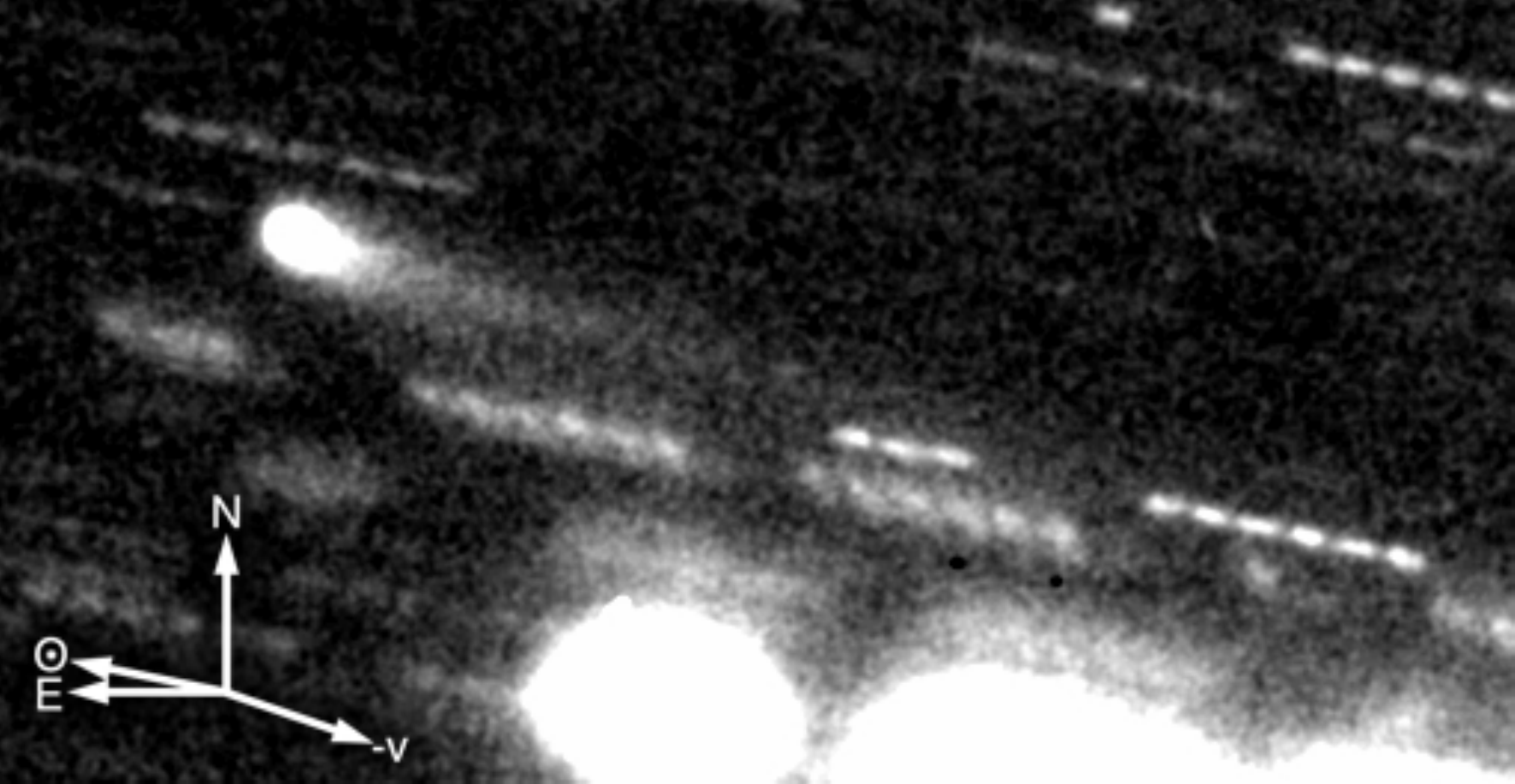}
  \caption{\small 238P imaged on UT 2005 November 10.  From \textit{Hsieh and Jewitt} (2006).\label{238P}}  
 \end{figure}

The second active asteroid in the main belt, 238P, exhibited a strong coma and dust tail when discovered in October 2005 at a heliocentric distance of $R\sim2.4$~AU (Figure \ref{238P}).  Like 259P, the nucleus of 238P is tiny, with a diameter of $\sim$0.8 km (\textit{Hsieh et al.}, 2011b).  It was observed to be active in both 2005 and 2010, with a period of inactivity in between, with a coma dust mass of order 10$^5$ kg and a production rate estimated (from published photometry) near $\sim$0.1 kg s$^{-1}$.  Also like 259P, 238P is dynamically unstable, with a survival time on the order of 20 Myr, although unlike 259P which may have been recently implanted at its current location, 238P is hypothesized to have diffused in eccentricity from its original location within the Themis family (which also contains 133P, 176P, and 288P) due to its proximity to the 2:1 mean-motion resonance with Jupiter (Haghighipour 2009).

\subsection{P/2012 T1 (PANSTARRS), $T_J$ =  3.134}

P/2012 T1 was discovered to exhibit a diffuse coma and a featureless fan-shaped antisolar tail in October 2012.  Photometric monitoring showed that the total scattering cross-section of the comet's coma and tail doubled over a period of about a month, remained approximately constant for another 1.5 months, and then declined by $\sim$60\% over the next 1.5 months (\textit{Hsieh et al.}, 2013).  Spectroscopic observations using the Keck I telescope found an upper limit CN production rate of $Q_{\rm CN}<1.5\times10^{23}$~mol~s$^{-1}$, and no evidence of absorption at 0.7~$\mu$m that would indicate the presence of hydrated minerals (\textit{Hsieh et al.}, 2013), while \textit{Herschel Space Telescope} observations were used to set an upper limit H$_2$O production rate of $Q_{\rm H_2O}<7.6\times10^{25}$~mol~s$^{-1}$ (\textit{O'Rourke et al.}, 2013).  Dust modeling by \textit{Moreno et al.}\ (2013) indicated that dust production began near perihelion and lasted for a period of $\sim4-6$ months, with a total ejected dust mass on the order of $10^7$~kg for maximum grain sizes of $a=1-10$~cm.

\subsection{313P/Gibbs (2014 S4), $T_J$ =  3.132}

\begin{figure}[h]
\epsscale{0.90}
\plotone{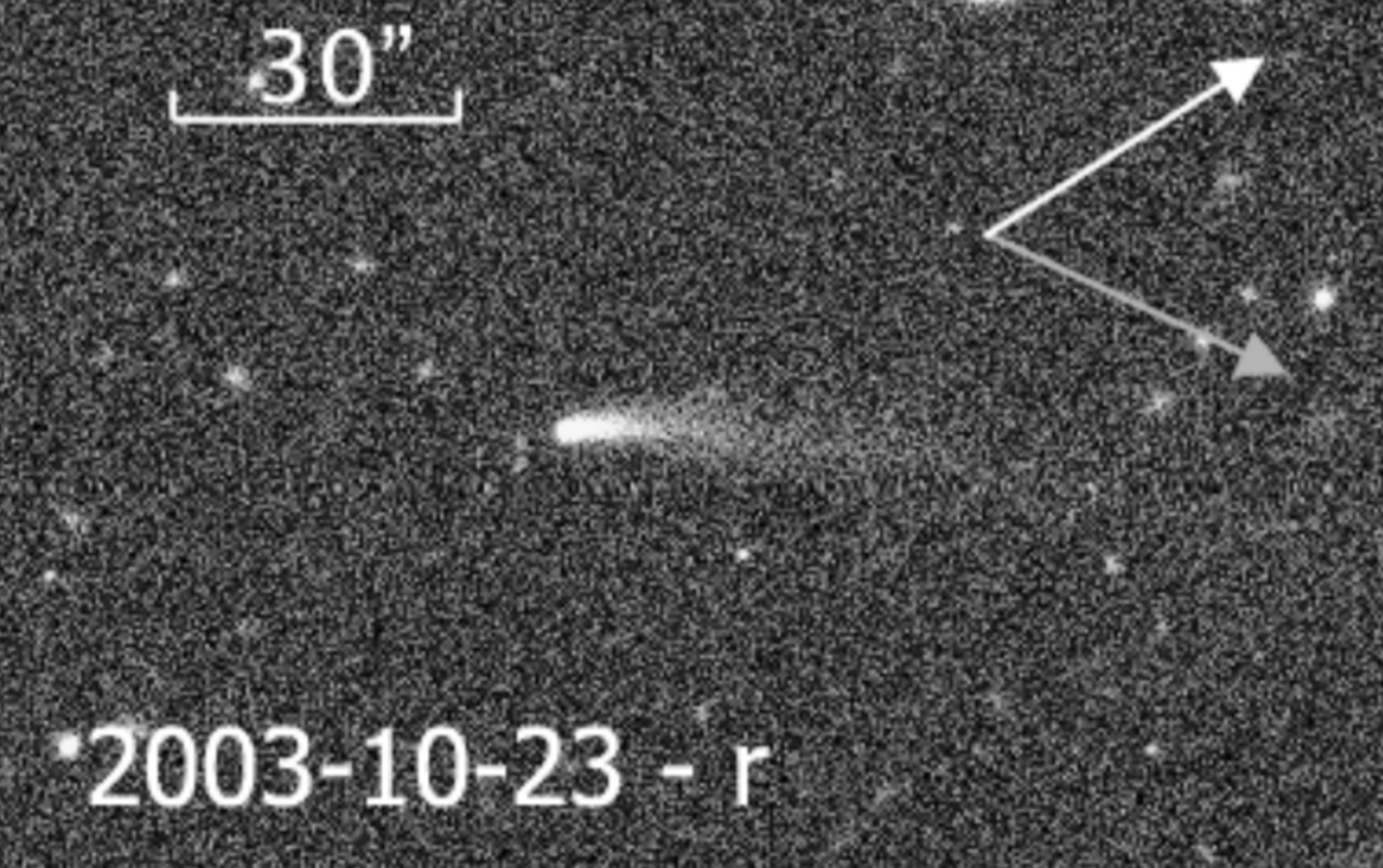}
\caption{\small Prediscovery image of 313P when at 2.47 AU on UT 2003 October 23.  White and grey arrows show the directions of the projected anti-solar and heliocentric velocity vectors, respectively.  From \textit{Hui and Jewitt} (2015).\label{313P}}  
\end{figure}

Discovered UT 2014 September 24 this object has semimajor axis 3.156 AU and is located near at least eight other active asteroids in the outer belt (Figure \ref{ae}).  The nucleus radius is 500 m (albedo 0.04 assumed). Prediscovery observations from 2003 (Figure \ref{313P}) reveal 313P as only the third object, after 133P and 238P, to be active in more than one orbit, consistent with mass loss driven by the sublimation of ice.  It displays a fan-shaped dust tail in both 2003 and 2014 that is well approximated by syndyne dust emission models, indicating that the dust is ejected over a period of at least 3 months during each active episode (\textit{Jewitt et al.}, 2015, \textit{Hsieh et al.}, 2015b, \textit{Hui and Jewitt}, 2015).   The object is found near two three-body mean-motion resonances with Jupiter and Saturn (11J-1S-5A and 10J+12S-7A), and its orbit has been found to be intrinsically chaotic with a Lyapunov time of $T_l=12\,000$~yr, yet numerical simulations show that it is stable over at least 50~Myr.  313P is the second active asteroid, after P/2012 T1, to be associated with the $\sim$155~Myr-old Lixiaohua asteroid family (\textit{Hsieh et al., 2015b}).

\subsection{P/2010 R2 (La Sagra), $T_J$ =  3.098}

The object was observed to be active from 2010 September to 2011 January, at $R$ = 2.6 to 2.7 AU, and has a measured nucleus size of $r=0.55\pm0.05$~km (\textit{Hsieh}, 2012c).   \textit{Moreno et al.,}~(2011b) inferred dust production at the peak rate of $\sim$4 kg s$^{-1}$, with centimeter-sized particles ejected at about 0.1 to 0.2 m s$^{-1}$.  A limit to the outgassing rate $Q_{CN} \le$3$\times$10$^{23}$ s$^{-1}$ (corresponding to $\sim$3 kg s$^{-1}$ in water) was placed spectroscopically (\textit{Hsieh et al.,}~2012c). The latter authors found (neglecting possible non-gravitational forces due to outgassing) that the orbit of P/2010 R2 is stable on timescales $\sim$100 Myr and argue that this object was likely formed in-situ.

\subsection{107P/(4015) Wilson-Harrington, $T_J$ =  3.083}

This object showed a prominent diffuse tail about 2$\arcmin$ in length on a blue sensitive photographic plate taken 1949 November 19, when at $R=1.148$~AU \textit{(Fernandez et al.,} (1997).   A red-sensitive plate taken nearly simultaneously shows only a hint of this tail.  107P appears strongly trailed on both plates owing to its non-sidereal motion.  The tail color ($B-R = -1$) is too blue to be caused by scattering from dust, and the tail position angle ($\sim15\degr$ from radial to the sun) is also inconsistent with the expected direction of a dust tail blown by radiation pressure.  107P was re-observed on November 22 and 25 but then showed no trace of a tail (\textit{Cunningham}, 1950), and no comet-like activity has been reported since (\textit{Chamberlin et al.}, 1996, \textit{Ishiguro et al.}, 2011c).  We regard 107P as the least convincing example of the active asteroids, both because the observations are old and unrepeated and because 107P lies very near the $T_J > 3.08$ cut-off.  \textit{Bottke et al.,} (2002) used a statistical dynamical model to conclude that there is a 4\% chance that 107P is a captured Jupiter family comet.

\subsection{Related Observations}
Features in the 3.0 to 3.5 $\mu$m reflection spectra of large main-belt asteroids (24) Themis (diameter 198$\pm$20 km) and (65) Cybele (273$\pm$12 km) have been interpreted as absorptions due to O-H stretch in water ice and C-H stretch in an unidentified organic molecule (\textit{Campins et al.,} 2010, \textit{Rivkin et al.,} 2010,  \textit{Licandro et al.,} 2011b).   The shape of the band in Themis requires a thin, widespread or global ice film (``frost'') only  100\AA~to 1000\AA~thick (\textit{Rivkin et al.,} 2010). However, such a thin film would be highly unstable to sublimation and lead to gas production rates 10$^5$ to 10$^6$ kg s$^{-1}$, violating observational limits set spectroscopically ($\lesssim$400 kg s$^{-1}$, \textit{Jewitt and Guilbert-Lepoutre} 2012).  The latter authors showed that, if water ice exists on these asteroids, it must be of high albedo ($>$0.3) and spatially confined to regions far from the subsolar point (e.g.~near the poles on objects having small obliquity) in order to keep sublimation below observational limits. The spectra of Themis and Cybele are similar to that of the iron-rich mineral goethite (\textit{Beck et al.,} 2011) but this offers a less plausible explanation because goethite is rare in meteorites and normally attributed to weathering of the meteorites once fallen on the Earth.

\section{\textbf{MECHANISMS}}
In this section we discuss possible mechanisms able to eject dust from asteroids.  The group properties offer few clues, given the modest size of the active asteroids sample (c.f.~Table \ref{physical}).  \textit{Licandro et al.,} (2011a), for example, suggest that the color distribution of active asteroids is different from that of classical cometary nuclei.  Even if future measurements prove this to be true, its significance is unclear given that a wide range of mechanisms are known to drive the observed activity in these bodies (\textit{Jewitt} 2012).  Furthermore, while they are discussed separately for clarity, it is likely that different mechanisms operate together in real objects.  (For example, the loss of particles produced in (3200) Phaethon by thermal disintegration is likely assisted by rotation and radiation pressure sweeping.  Sublimation driven mass loss from 133P is probably also rotation assisted).

\subsection{Rotational Mass Loss}

For a sphere of density $\rho$ the critical period at which the gravitational acceleration equals the centripetal acceleration at the equator is 

\begin{equation}
P_c = \left(\frac{3 \pi}{G \rho}\right)^{1/2}
\label{critical}
\end{equation}

\noindent where $G$ is the gravitational constant.  For example, with $\rho$ = 1000 kg m$^{-3}$, $P_c$ = 3.3 hr.  The critical period is independent of the asteroid size, is shorter for higher densities and longer for elongated spheroids in rotation about a minor axis (by a factor approximately equal to the ratio of the long to short axes of the body).  The rotation and the shape of a fluid body are related, in equilibrium, by the classic MacLaurin and Jacobi ellipsoid series.  However, while most asteroids may have been severely weakened by repeated impact fracturing and the formation of a rubble-pile structure, they are not strengthless. van der Waals and other weak forces can imbue a rubble pile with a cohesive strength while friction can provide resistance to deformation even in the absence of cohesion, and critical periods much shorter than given by Equation (\ref{critical}) are possible (\textit{Holsapple} 2007, \textit{Sanchez and Scheeres} 2014).   Calculation of the rotational stability of rubble-piles remains an important but  challenging and under-constrained problem in asteroid science.

Evidence for rotational instability comes from the observation that most asteroids larger than a few hundred meters in size rotate more slowly than  a ``barrier'' period at about 2.2 hr (\textit{Warner et al.,} 2009).    This is widely interpreted as meaning that faster rotating asteroids have lost mass or even disrupted due to centripetal forces.  Recent observations of active asteroids 311P and P/2013 R3 (and possibly P/2010 A2) appear to show asteroids losing mass rotationally, so that we can begin to study this process as it happens.

In principle, the rotation rates of asteroids can be driven to critical values by external torques exerted by the gravity of other objects, by chance impacts, by outgassing and by electromagnetic radiation.  In the main belt, gravitational torques are negligible (except within binary and other multiple systems). Collisions  add angular momentum stochastically, leading to a slow random walk towards larger angular momenta.  Outgassing from icy asteroids can be very efficient in changing the spin but, for ice-free asteroids, radiation torques offer the most potential for driving the rotation steadily up to critical values.

Power absorbed from the Sun by the surface of an asteroid is re-radiated  to space as heat.  Asteroids are aspherical and anisothermal, causing the radiated infrared photons to be emitted anisotropically.  The angle-averaged momentum carried by thermal photons per second corresponds to a net reaction thrust on the asteroid, known as the Yarkovsky force.  It has important dynamical consequences on small main-belt asteroids. If the vector representing the net force does not pass through the center of mass of the asteroid, the result is the so-called YORP torque, which can change the magnitude of the spin and excite precession (see the chapter by \textit{Vokrouhlicky et al.} in this volume).  

We focus on changes in the magnitude of the spin, and estimate the relevant timescale from the ratio of the rotational angular momentum, $L$, to the torque, $T$. Ignoring the vector nature of these quantities, for simplicity, the torque is proportional to the number of photons radiated per second, which varies in proportion to $r_n^2/R^2$, where $r_n$ is the asteroid radius and $R$ is the distance from the Sun.  Torque also depends on the moment-arm, defined as the perpendicular distance between the instantaneous direction of the net force and the center of mass. Statistically, at least, we expect the moment arm $\propto r_n$.  Together, these dependences give  $T \propto r_n^3/ R^2$. Meanwhile, the spin angular momentum  is  $L \propto M r_n^2 \omega$, where $M$ is the body mass and $\omega$ the angular rotation rate, related to the rotational period, $P$, by $\omega = 2\pi/P$.  Substituting $M \propto \rho r_n^3$ gives $L \propto \rho r_n^5 \omega$. Finally, the timescale for YORP to change the angular momentum is $\tau_y = L/T$, or $\tau_y= K \rho r_n^2 R^2 \omega$, where $K$ is a constant.  

The value of constant $K$ depends sensitively on the body shape, surface texture, thermal properties and spin vector of the asteroid.  We estimate its magnitude from measurements of YORP acceleration in five asteroids (\textit{Rozitis and Green} 2013, \textit{Lowry et al.,} 2014), scaled to assumed density $\rho$ = 2000 kg m$^{-3}$ and spin period $P$ = 5 hr, to obtain

\begin{equation}
\tau_{y} (Myr) \sim 4  \left[\frac{r_n}{1~km}\right]^2  \left[\frac{R}{3~AU}\right]^2.
\label{yorp}
\end{equation}

Order of magnitude deviations from Equation (\ref{yorp})  can result from peculiarities of the asteroid shape, texture, thermal properties and spin.  Still, Equation (\ref{yorp}) gives a crude but useful estimate, by showing that the angular momentum of a 1 km radius asteroid at 3 AU can be modified significantly on timescales of just a few Myr.  A sustained YORP torque can drive a kilometer-scale body to rotational instability in a few times $\tau_y$, and the spins of sub-kilometer, low-density, near-Sun asteroids should be particularly susceptible to YORP torques. Setting $\tau_y <$ 4.5$\times$10$^9$ yr in Equation (\ref{yorp}) shows that asteroids as large as $r_n \sim$ 30km can be affected by YORP in the age of the solar system, provided they survive against collisions for such a long time.  Most of the objects in Table (\ref{physical}) are small enough to be potentially affected by YORP.

\begin{figure}[h]
 \epsscale{0.95}
\plotone{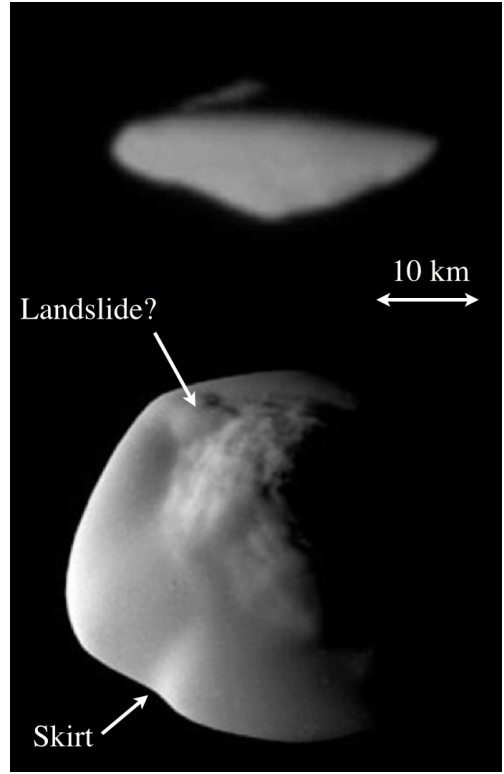}
 \caption{\small Two views of UFO-shaped Saturnian satellite Atlas showing a dust skirt around the equator where the effective gravitational attraction to the body approaches zero.   Atlas is a slow-rotator (period 14 hr), but it fills its Roche lobe within the Saturninan system, making it a suitable analog for a body at the stability limit (\textit{Minton} 2008). The paucity of craters indicates youth, which results from the mobility of surface dust. A possible landslide is visible on the dust skirt; corresponding landslides or avalanches leading to escape  are likely responsible for the impulsive tail ejections in 311P (c.f.~Figure (\ref{311P})).  The bulk density $\sim$400 kg m$^{-3}$.  Credit: NASA/JPL/SSI.   \label{atlas}}  
 \end{figure}

A major complication is that YORP torques can change not just the rotation state but the shape of an asteroid, as material slides, bounces or rolls towards the rotational equator on its way to escape (e.g.~\textit{Harris et al.,} 2009, \textit{Statler} 2009).     Changing the shape affects the magnitude and possibly the direction of the YORP torque (\textit{Cotto-Figueroa et al.,} 2014), creating a feedback loop that can change the timescale for spin-up relative to the value in Equation (\ref{yorp}).  Evidence for equatorial accumulation of particulate matter is present in images of UFO-shaped Saturnian satellites (c.f.~Figure \ref{atlas}) and in the shapes of some rapidly-rotating near-Earth asteroids, as determined by radar (\textit{Ostro et al.,} 2006).

\begin{figure}[h]
 \epsscale{1.0}
\plotone{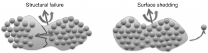}
 \caption{\small Schematic diagram distinguishing (left panel)  structural failure, as may be seen in P/2013 R3, from (right panel) surface shedding (for which 311P is a candidate).  From \textit{Hirabayashi and Scheeres,} (2014). \label{hirabayashi}}  
 \end{figure}

In part because of the complex interplay between torque, spin, material properties, shape and mass loss, no comprehensive models of asteroid evolution and disruption by rotation have been computed, although many interesting approximations exist (\textit{Holsapple} 2007, 2010,  \textit{Walsh et al.,} 2008, 2012, \textit{Jacobson and Scheeres} 2011,  \textit{Marzari et al.,} 2011, \textit{Hirabayashi et al.,} 2014a,  \textit{Sanchez and Scheeres} 2012, 2014, \textit{Cotto-Figueroa et al.,} 2014, \textit{Scheeres} 2015).  These models, not always in agreement with one another, attempt to investigate the role of small torques in  the production of distinctive body spins and shapes, in mass-shedding and structural failure (c.f.~Figure \ref{hirabayashi}), and in the formation of both binaries and unbound pairs (\textit{Pravec et al.,} 2010, \textit{Moskovitz} 2012, \textit{Polishook} 2014).  

The cohesive strength is a key parameter in all such models because, given sufficient cohesive strength, an asteroid can resist rotational forces at an arbitrarily high rotation rate (e.g.~\textit{Rozitis et al.,} 2014).   Fortunately, new observations of active asteroids allow us to obtain a useful estimate of the cohesive strength.  

First, we consider a toy model of two bodies in contact, both of density $\rho$ and having dimensions $r_p$ and $r_s$, with $r_p \gg r_s$.    Assume that the smaller body sits on the rotational equator of the larger body and take the area of contact between them to be $\sim r_s^2$.  Let the critical frequency of rotation at which gravitational attraction is balanced by centripetal acceleration be $\omega_c$.  Then, when rotating with an angular frequency $\omega > \omega_c$,  the minimum cohesive strength needed to bind the two bodies together may be written

\begin{equation}
S \sim \frac{m_s r_p (\omega^2 - \omega_c^2)}{ r_s^2}
\label{escape}
\end{equation}

\noindent  where $m_s$ is the mass of the smaller component.  If detachment occurs when  $\omega^2 >>  \omega_c^2$, the smaller mass $m_s$ will leave with a relative velocity, $\Delta V$, comparable to the instantaneous equatorial velocity of the primary, namely $\Delta V = r_p \omega$.  Substituting for $\Delta V$  and $m_s = \rho r_s^3$ into Equation (\ref{escape}) we obtain

\begin{equation}
S \sim \rho \left(\frac{r_s}{r_p}\right) (\Delta V)^2
\label{strength}
\end{equation}

\noindent The six measured components of P/2013 R3  (Table 2 of \textit{Jewitt et al.,} 2014b) are equivalent in volume to a single precursor body of dimension $r_p$ = 350 m. Four of these six nuclei have $r_s \sim$ 200 m and are separating at characteristic speeds 0.2 $\lesssim \Delta V \lesssim$ 0.5 m s$^{-1}$.  Assuming density $\rho$ = 1000 kg m$^{-3}$  Equation (\ref{strength}) gives an estimate of the cohesive strength as 16 $\lesssim S \lesssim$ 140 N m$^{-2}$.  
A mathematically more refined analysis, in which the initial body shape is represented by a rotational ellipsoid, gives 40 $\lesssim S \lesssim$ 210 N m$^{-2}$ (\textit{Hirabayashi et al.,} 2014b).  In both cases, these very small cohesive strengths are comparable to the $\sim$25 N m$^{-2}$ strength expected of a  rubble pile bound by van der Waals forces acting on the smallest particles (\textit{Sanchez and Scheeres} 2014).  The small size of the P/2013 R3 precursor body is compatible with a short YORP timescale ($\tau_y \sim$ 0.5 Myr by Equation \ref{yorp}), lending credibility to the YORP spin-up hypothesis for this object.  Active asteroid 311P has also been interpreted (\textit{Jewitt et al.,} 2013c,2014c) as rotational instability of a different sort, in which only surface particulates escape from the rotating central body (\textit{Hirabayashi et al.,} 2014).  

 \begin{figure}[h]
 \epsscale{0.99}
\plotone{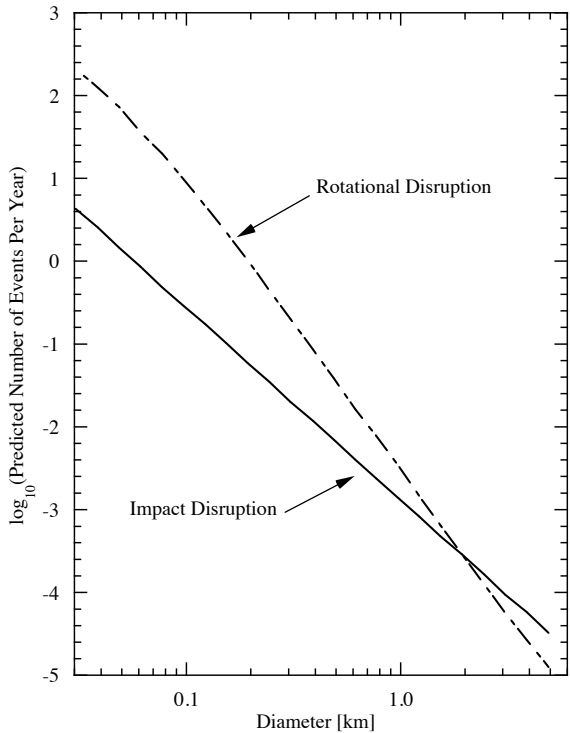}
  \caption{\small Model rates of asteroid disruption due to impact  and rotational bursting as a function of asteroid diameter.  From \textit{Marzari et al.,} (2011). \label{marzari}}  
 \end{figure}

This first determination of the cohesive strength, if broadly applicable, shows that rotation of  asteroids can play a major disruptive role.  Indeed, the rotational disruption rate may exceed the impact destruction rate for asteroids $\lesssim$1 km in diameter (Figure \ref{marzari}).  However, many questions remain about the process.   In 311P, what determines the intervals between successive dust releases and for how long will they continue?  More specifically, does the distribution of dust releases resemble the power law spectra of self-organized critical sandpiles (\textit{Laurson et al.,} 2005), as might be expected if avalanches are responsible?  Could this object be in the process of forming a binary (c.f.~\textit{Walsh et al.,} 2008, \textit{Jacobson and Scheeres} 2011) or has it already formed one?  Will the nucleus eventually be driven to structural instability, as has occurred in P/2013 R3?  Or could  311P  be a fragment of a rotationally disrupted precursor body whose components are so widely spread as to have escaped notice?

\subsection{Impacts}
\label{impacts}

The average velocity dispersion among main-belt asteroids is $U \sim$ 5 km s$^{-1}$, enough to cause vaporization as well as shattering and, potentially, disruption upon impact.    Experiments at laboratory scales show that the ratio of $m_e$, the mass ejected faster than speed $v$,  to the projectile mass, $M$, is 

\begin{equation}
\frac{m_e}{M} = A \left(\frac{v}{U}\right)^{\alpha}
\label{ejecta}
\end{equation}

\noindent with $A \sim$ 0.01 and, very roughly, $\alpha \sim$ -1.5 (\textit{Housen and Holsapple} 2011).  For an assumed spherical projectile, the mass is $M = 4\pi \rho r_p^3/3$, where $r_p$ is the radius. Equation (\ref{ejecta}) is valid provided the impact does not have enough energy to completely disrupt the target asteroid.  For example, an impact at 5 km s$^{-1}$ into a 1 km radius body, with escape velocity $v_e \sim$ 1 m s$^{-1}$, would have an ejecta to projectile mass ratio $m_e/M \sim$ 3500 by Equation (\ref{ejecta}).  The mass of ejecta and the scattering cross-section, $C$, are related by

\begin{equation}
m_e = \chi \rho \overline{a} C
\label{cross-section}
\end{equation}

\noindent where $\overline{a}$ is the weighted mean particle size in the ejecta and $\chi$ is a dimensionless constant of order unity.  Very modest projectiles are capable of creating ejecta with a substantial cross-section.   We write  $C = f \pi r_n^2$, where $r_n$ is the radius of the target asteroid and $f$ is a constant.  The case $f$ = 1 corresponds to ejecta with a cross-section equal to that of the target asteroid.  Such an impact would result in a doubling of the asteroid brightness, and therefore should be readily detected.  We set $v$ = $(8\pi G \rho/3)^{1/2} r_n$, the gravitational escape speed from the target asteroid, and combine Equations (\ref{ejecta}) and (\ref{cross-section}) to obtain

\begin{equation}
r_p = \left(\frac{3 \overline{a} f}{4 A}\right)^{1/3} \left(\frac{8\pi G \rho}{3}\right)^{-\alpha/6} U^{\alpha/3} r_n^{((2-\alpha)/3)}
\label{brighten}
\end{equation} 

\noindent  Equation (\ref{brighten}) assumes that the densities of the target and projectile are equal.  Substituting the above parameters, Equation (\ref{brighten}) reduces to 

\begin{equation}
r_p = 0.29 f^{1/3}  r_n^{7/6}
\label{brighten2}
\end{equation}

\begin{figure}[h]
 \epsscale{1.0}
\plotone{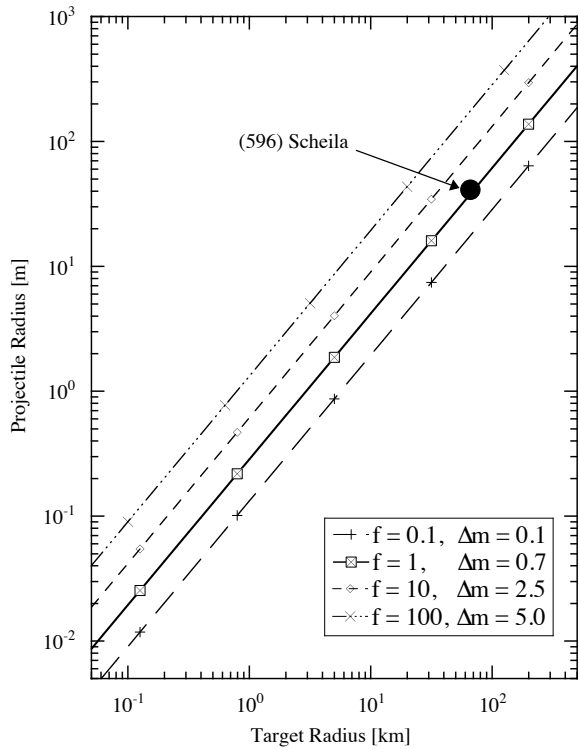}
 \caption{\small Projectile radius (y-axis) needed to generate ejecta with cross-section  $f$ = 0.1, 1, 10 and 100 times the geometric cross-section of the target asteroid, whose radius is plotted on the x-axis (c.f.~Equation \ref{brighten2}).  Impacted asteroid (596) Scheila ($r_p$ = 65 km) brightened by about 1 magnitude following impact.  The figure shows that the projectile radius was several 10's of meters.   \label{dm_plot}}  
 \end{figure}

\noindent Equation (\ref{brighten2}) is plotted in Figure (\ref{dm_plot}) for $f$ = 0.1, 1, 10 and 100, corresponding to changes in the apparent magnitude due to impact of $\Delta m \sim 2.5 \log_{10} f$ = 0.1, 0.7, 2.5 and 5 magnitudes, respectively.  The Figure shows that a 20 cm projectile striking a 1 kilometer asteroid would eject enough material to double the total cross-section and brightness.  Very modest impacts should create observable ejecta signatures.

The number of boulders in the asteroid belt is very uncertain, because these small objects are too faint to be observed directly from Earth.  Models extrapolating from larger sizes suggest that the number of objects with $r_p >$ 1 m  is roughly 10$^{13\pm1}$ (\textit{O'Brien and Greenberg,} 2005).    The rates at which small-scale impacts occur is correspondingly uncertain by at least an order of magnitude. 

Survey observations by \textit{Denneau et al.,} (2014) have been interpreted, through a simple model,  to show a discrepancy with the rate of asteroid disruption predicted by \textit{Bottke et al.,} (2005).  

\subsection{Thermal Disintegration}
Forces caused by thermal expansion can exceed the fracture strength of a material, leading to cracking and the production of dust.  On a low gravity body, excess thermal strain energy can eject fragments faster than the gravitational escape speed, leading to a net loss of material (\textit{Jewitt} 2012).    In a uniform material, stresses arise from differential thermal expansion across a temperature gradient.  Most minerals are non-uniform in structure, consisting of compositionally distinct grains.  Non-uniform materials are susceptible to fracture because of differential expansion between the component materials, with or without a temperature gradient.     The thermal skin depth in the material (across which temperature gradients are particularly marked, and variable) is related to the period of the temperature variation, $\tau$, and to the thermal diffusivity, $\kappa$, by $\ell \sim (\kappa \tau)^{1/2}$. For example, an asteroid with diffusivity $\kappa$ = 10$^{-6}$ m$^2$ s$^{-1}$, rotating with period $\tau$ = 5 hr would have a skin depth  $\ell \sim$ 0.1 m.   Thermal fracture occurring within a thermal skin depth of the physical surface may contribute to the production of regolith (\textit{Delbo et al.,} 2014).  Desiccation (the loss of bound water from a hydrated mineral) is a separate process leading to fracture and dust production when the stresses induced by shrinkage exceed the material strength.   On Earth, mudcracks in dry lake beds made of clay minerals are the most easily recognized examples of desiccation cracking.

Among the active asteroids, the most convincing evidence for thermal disintegration is from (3200) Phaethon (\textit{Li and Jewitt} 2013, \textit{Jewitt and Li} 2010, 2013).   The eccentric orbit and high obliquity (Phaethon's pole is at ecliptic coordinates $\lambda, \beta$ = 85\degr$\pm$14\degr, -20\degr$\pm$10\degr, Ansdell et al., 2014) lead to highly non-uniform surface heating, with peak temperatures near 1000 K at perihelion (\textit{Ohtsuka et al.,} 2009).   The temperature is further modulated by Phaethon's rapid rotation (3.6 hr period), with a diurnal variation of hundreds of Kelvin. Furthermore, the escape speed from the $\sim$5 km diameter nucleus is a few meters per second, sufficiently small that grains produced by fracture can easily be ejected (\textit{Jewitt} 2012).  While Phaethon's composition is unknown, hydrated mineralogies (montmorillonite clays) have been suggested based on the blue optical/near-infrared reflection spectrum (\textit{Licandro et al.,} 2007).  Hydrated materials, if present, would be highly susceptible to desiccation, shrinkage and fracture.  Nine asteroids with  perihelia $\le$0.25 AU (subsolar temperatures $\ge$800 K) have been searched for activity, setting upper limits to the mass loss rate from $\le$ 0.1 to 1 kg s$^{-1}$ (\textit{Jewitt} 2013).

On a larger scale, thermal disintegration may be important in dust disks surrounding white dwarf stars.  Such disks are inferred from photospheric, heavy-element pollution (\textit{Jura and Xu} 2013)

\subsection{Sublimation of Ice}
\label{ice}
\begin{figure}[h]
 \epsscale{0.99}
\plotone{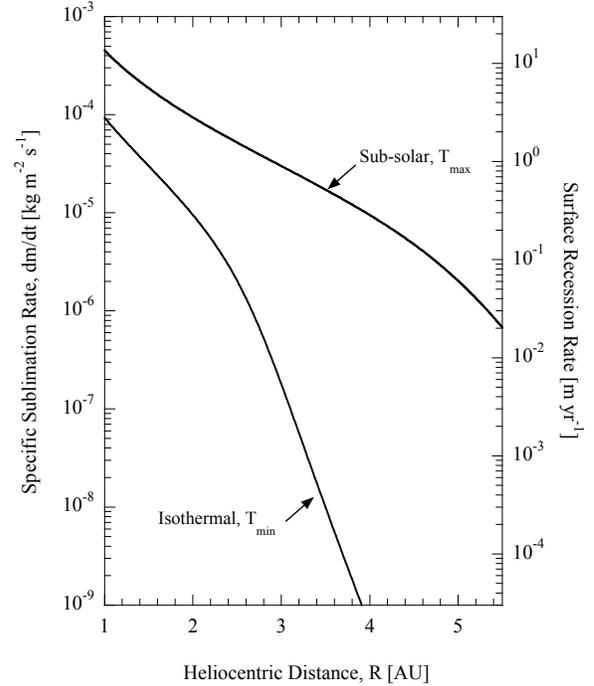}
\caption{\small Solutions to Equation (\ref{sublimation}) for the extreme cases of subsolar sublimation ($\theta$ = 0\degr) and for isothermal sublimation. The  right hand axis shows the rate of recession of the sublimating surface due to mass loss, given by $dr/dt = \rho^{-1} dm/dt$ (m s$^{-1}$), where $\rho$ is the density, here taken to be 1000 kg m$^{-3}$. From \textit{Jewitt} (2012). 
\label{dmbdt}}  
\end{figure}

Ice exposed at the surface of an asteroid at distance $R$ AU from the Sun will sublimate, in equilibrium, at a specific rate $dm/dt$ (kg m$^{-2}$ s$^{-1}$), given by the solution to

\begin{equation}
\frac{F_{\odot} (1 - A)}{R^2} cos(\theta) = \epsilon \sigma T^4 + L(T) \frac{dm}{dt}
\label{sublimation}
\end{equation}

\noindent where the term on the left represents the absorbed solar power per unit area, the first term on the right represents thermal radiation to space per unit area and the second term represents the power used per unit area in breaking bonds to sublimate ice.  $F_{\odot}$ = 1360 W m$^{-2}$ is the solar constant, $A$ is the Bond albedo, $\epsilon$ is the emissivity of the asteroid and $L(T)$ is the latent heat of sublimation of ice at temperature $T$.  Here $\theta$ is the angle between the Sun and the surface normal as seen from the sublimating surface.   Values of $cos(\theta)$ range from 1/4 (the isothermal case, corresponding to the lowest possible temperatures) to 1 (for normal illumination at the sub-solar point, the maximum temperature case).  Conduction is ignored for simplicity - its effect will be to reduce $dm/dt$ relative to the value computed from Equation (\ref{sublimation}) by an amount that depends on the conductivity (measurements of small body regoliths suggest that the conductivity is very small, so that the error incurred by its neglect should be minor).  Equation (\ref{sublimation}) can be solved in combination with the Clausius-Clapeyron relation for water ice, to evaluate $dm/dt(R)$, as in Figure (\ref{dmbdt}).  The Figure also shows (on the right hand axis) the rate of recession of the sublimating surface due to mass loss, given by $dr/dt = \rho^{-1} dm/dt$ (m s$^{-1}$). 

Maximum sublimation rates vary by nearly an order of magnitude from $\sim$10$^{-4}$ kg m$^{-2}$ s$^{-1}$ at 2 AU to $\sim$10$^{-5}$ kg m$^{-2}$ s$^{-1}$ at 3.5 AU, corresponding to the inner and outer edges of the main -belt.  The corresponding surface recession rates are from a few meters to a few$\times$0.1 meters per year.  Water ice is thermodynamically unstable when exposed to sunlight at asteroid belt distances, but can be preserved on billion-year timescales beneath meter-thick or greater layers of porous refractory debris (\textit{Fanale and Salvail} 1989, \textit{Schorghofer} 2008).  Sublimation at rates sufficient to launch dust from a small body would then require that the protective regolith be removed, perhaps by a small impact or some other disturbance of the surface.  In this scenario, sublimation would proceed from an exposed region until a new refractory covering, consisting of particles too large to be lifted by gas drag (a so-called ``rubble mantle''), chokes off the flow.   The ratio of the lifetime of the sublimating surface patch to the interval between successive impacts (or other surface disturbances) defines the ``duty cycle'', $f_d$.  For example, the timescale for formation of a rubble mantle on a kilometer-scale body at 3 AU is estimated from a simple model (Figure 4 of \textit{Jewitt} 2002) at $\sim$1 yr, increasing to 10 yr at 4 AU.  The interval between impacts of meter-scale projectiles (needed to expose ice sufficient to account for the mass loss) onto this body is perhaps $t \sim$ 10$^4$ yr, giving $f_d \sim$ 10$^{-4}$ to 10$^{-3}$.   Neither the re-sealing timescale nor the excavation timescale can be specified with confidence, but $f_d \ll$ 1 is assured.  As noted in Section \ref{impacts}, impacts at 5 km s$^{-1}$ are highly erosive, with an ejecta to projectile mass ratio $m_e/M \sim$ 3500 for a 1 km target asteroid (Equation \ref{ejecta}).  At this rate, the asteroid can survive for $\sim$10$^9$ yr.

\subsection{Radiation Pressure Sweeping}
Particles lifted above the surface of an asteroid are susceptible to solar radiation pressure and, if sufficiently small,  can be blown away.  The critical size for radiation pressure sweeping on a non-rotating asteroid is (\textit{Jewitt} 2012)

\begin{equation}
a_{\beta}(\mu m)= 10 \left(\frac{1~km}{r_n}\right) \left( \frac{1~AU}{R}\right)^2.
\end{equation}

\noindent At $R$ = 3 AU, a 1 $\mu$m radius grain detached from the surface (by any process) can be blown away from a non-rotating asteroid 1 km in radius.

\subsection{Electrostatics and Gardening}
Dielectric surfaces exposed to ionizing (UV and X-ray) solar radiation develop a positive electric potential (of perhaps 5 to 10 V) through the  loss of photoelectrons. Conversely, electrons impacting in shadowed regions imbue a locally negative potential. The resulting electric fields near shadow boundaries can grow large enough ($\sim$10 V m$^{-1}$ to 100 V m$^{-1}$) to mobilize dust.  On the Moon, horizon glow (Rennilson and Criswell 1974) and dust impact counter experiments (\textit{Berg et al.,} 1976) show that charging effects near the terminator  lift 10 $\mu$m sized dust particles to meter heights and greater, at implied ejection speeds, $\sim$1 m s$^{-1}$.  Smooth dust ponds on asteroids strongly suggest electrostatic  mobilization (but not ejection)  of grains in a much lower gravity environment (\textit{Poppe et al.,} 2012, see also the chapter by \textit{Murdoch et al.,} in this volume).  

On small asteroids, electrostatic forces can potentially accelerate particles to speeds exceeding the gravitational escape speed.  To within a numerical factor of order unity, particles smaller than a critical radius

\begin{equation}
a_e (\mu m) \sim 2 \left(\frac{1\ km}{r_n}\right)
\label{electrostatic}
\end{equation}

\noindent can be electrostatically ejected against gravity from a non-rotating body of radius $r_n$ (\textit{Jewitt} 2012).  

Larger particles are too heavy to lift against gravity, at least on a non-rotating body, whereas smaller particles are trapped by cohesive forces on the surface  (\textit{Hartzell and Scheeres} 2011).  Electrostatically ejected particles should therefore be confined to a narrow range of sizes near $a_e$.  Note that Equation (\ref{electrostatic}) is independent of the heliocentric distance because the potential, $V$, depends on photon energy, not flux.  Since, in the asteroid belt at 2 to 3 AU, $a_e \lesssim a_{\beta}$, particles launched electrostatically should be picked up by solar radiation pressure and propelled into comet-like dust tails.

Although electrostatic  ejection must operate at some level,  in practice the rate of loss of small particles  from an asteroid will be limited by the  rate of their production, presumably by micrometeorite bombardment.  This distinguishes the role of electrostatics on the asteroids from the circumstance on the Moon, where ejected particles fall back to the surface under the high lunar gravity, to be repeatedly relaunched at each terminator crossing (\textit{Berg et al.,} 1976).  

We crudely limit the  ejected mass flux from asteroids as follows.  On the Moon, impact overturn of the regolith (``gardening'') reaches to depths $\delta r \lesssim$1 m on timescale $t$ = 1 Gyr (\textit{Gault et al.,} 1974).  Impact-produced fragments that are largely retained on the Moon by gravity will, on the small asteroids, be immediately lost.  Therefore, a strong upper limit to the electrostatic mass loss  from asteroids can be obtained by assuming that  the gardening rates are comparable and that none of the gardened layer is retained. The mass loss rate from a spherical asteroid is $dM/dt \sim 4 \pi r_n^2 \rho \delta r/ t$, where $r_n$ is the radius and $\rho$ is the density.  For an $r_n$ = 1 km asteroid with $\rho$ = 2000 kg m$^{-3}$ and $\delta r$ = 1 m, this gives $dM/dt \lesssim  10^{-6}$ kg s$^{-1}$.  This is 4 to 6 orders of magnitude smaller than the mass loss rates inferred from observations of active asteroids (Table \ref{physical}).  Of course, we cannot exclude spatially or temporally enhanced micrometeorite impact fluxes in the asteroid belt, but enhancements sufficient to bring the mass loss rates up to those in Table (\ref{physical}) seem unlikely.  The only observation possibly supporting the existence of widespread, very low level mass loss is by \textit{Sonnett et al.,} (2011).  If this observation is confirmed, it might be evidence for very low level ejection of dust owing to gardening losses, possibly assisted by electrostatic effects.

\subsection{Observational Diagnostics} 

Diagnosing the cause of mass loss from any particular object is difficult, both because of limitations of the data and because of uncertainties in models of the mass loss mechanisms.  It is often easier to rule out possible explanations than to rule them in.  Spectroscopic detections of water vapor or another sublimation product would provide the most solid evidence of sublimation-driven activity, but such evidence has only been obtained for Ceres (\textit{K{\"u}ppers et al.}, 2014).  Unsuccessful attempts have been made to detect water vapor or CN emission from several other active asteroids (e.g., \textit{Jewitt et al.}, 2009; \textit{Licandro et al.}, 2011a, 2013; \textit{Hsieh et al.}, 2012a,b,c, 2013; \textit{de Val-Borro et al.}, 2012; \textit{O'Rourke et al.}, 2013).  Typical derived upper limits to gas loss rates are $\lesssim$1 kg s$^{-1}$. Unfortunately, the significance of these non-detections is inconclusive, for two reasons.  First, CN  is a prominent trace species in the spectra of Kuiper belt and Oort cloud comets (with H$_2$O/CN $\sim$360); it might be strongly depleted in icy asteroids.  Second, spectra are typically obtained long after continuum observations have revealed the mass loss - the gas could simply have escaped on a timescale short compared to the radiation pressure sweeping time for dust. 

Another potential indicator is the recoil (or ``rocket'') acceleration produced on the nucleus by anisotropic sublimation.  Specifically, a non-gravitational acceleration, $\alpha$, implies a mass loss rate

\begin{equation}
\frac{dM}{dt} = \frac{\alpha M_n}{f_{\theta}v_{0}}
\label{rocket}
\end{equation}


\noindent where $M_n$  is the mass of the nucleus and $v_{0}$ is the velocity of the material ejected.  Dimensionless parameter $f_{\theta}$ accounts for the directional pattern of the mass loss, with $f_{\theta}$ = 0 and 1 corresponding to isotropic and perfectly collimated ejection, respectively.    In fact, non-gravitational acceleration of 133P at $\alpha$ = 2.5$\times$10$^{-10}$ m s$^{-2}$, has been reported (\textit{Chesley et al.}, 2010), albeit with only 3$\sigma$ statistical significance. To obtain the smallest possible estimate of $dM/dt$ from Equation (\ref{rocket}), we maximize $f_{\theta}$ = 1 and use the sound speed in H$_2$O  gas at the blackbody temperature appropriate to 3 AU, namely $v_{0}$ = 500 m s$^{-1}$.  Approximated as a sphere of radius 2.2 km and density 1000 kg m$^{-3}$, the nucleus mass is $M_n$ = 4$\times$10$^{13}$ kg, giving  $dM/dt \sim$ 20 kg s$^{-1}$.  This rate exceeds the dust production rates measured  in 133P by two to three orders of magnitude, forcing us to conclude either that $\alpha$ has another cause or that the reported value should be considered only as an upper limit.  \textit{Nugent et al.}~(2012) measured small semi-major axis drift rates in 42 near-Earth asteroids consistent with the Yarkovsky (radiation) acceleration.  An additional 12 asteroids showed drift rates too large to be explained by the Yarkovsky effect.   If confirmed, the accelerations of these asteroids, none of which is a known active asteroid, might suggest anisotropic mass loss.

\begin{figure}[h]
 \epsscale{0.99}
\plotone{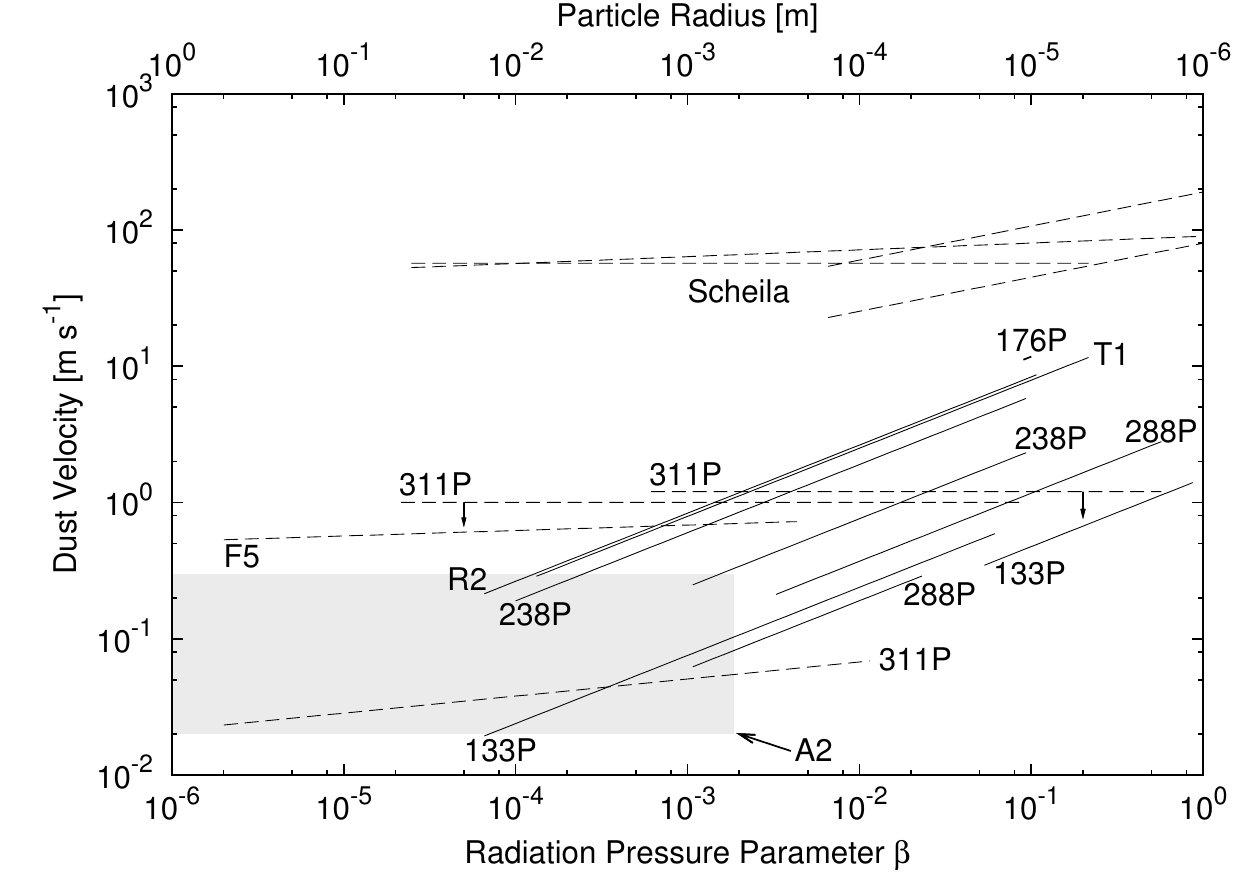}
  \caption{\small Dust ejection velocities as function of the radiation pressure coefficient $\beta$ or particle size. The plot is a compilation of published models for individual objects. Such models typically simulate the motion of dust grains ejected from the nucleus over a range of times, particle sizes, and velocities. The best-fitting parameters are determined from comparison with the observed shapes of the dust tails.
Solid lines indicate objects with protracted, likely sublimation-driven dust emission, dashed lines and the shaded area for P/2010 A2 refer to objects with impulsive emission. Arrows indicate upper limits, the shaded area for P/2010 A2 illustrates that particles of all sizes were ejected with the same range of velocities.  \label{V_vs_beta}}  
 \end{figure}

Less direct inferences can be made in the absence of spectroscopic detection of gas.  Activity which repeats on the orbital timescale, as in 133P and 238P, is naturally explained only by sublimation.   A prolonged period of dust emission is also suggestive of sublimation and difficult to explain as the result of a simple impact.    However, prolonged emission may not be unique to sublimation.  Evidence from 311P shows that mass shedding can continue for months, raising the potential for confusion with protracted emission caused by sublimation.  The size vs.~velocity relation for ejected dust particles also differs between mechanisms, with gas drag giving $v \propto a^{-1/2}$, where $a$ is the dust radius, and rotational instabilities and impact giving a flatter dependence (Figure \ref{V_vs_beta}).

Sublimation will preferentially lift small, micron-sized particles that are quickly dispersed by radiation pressure. Objects typically have a bright coma and tail for a few months, as long as the sublimation is on-going, followed by a rapid fading and return to the appearance of a point source. During rotational breakup, the ejecta sizes are not constrained at all, due to the compensation of gravity by the centrifugal force. Also impacts are expected to excavate much larger pieces of debris, depending on the escape speed from the parent body.
Characteristically, small objects suspected to have undergone an impact or rotational breakup display long-lived trails of cm- to m-sized debris along their orbits for many years after the event, because such large grains are not efficiently removed by radiation pressure. (596) Scheila, though clearly impacted, did not develop a debris trail because the high escape speed prevented the ejection of large debris.
The absence of debris trails in sublimation-driven active asteroids is different from comets, likely because cometary sublimation is much stronger and able to lift also large debris.

Numerical modeling is necessary because the appearance of long-lived activity can be produced either by the lingering of large, slow-dissipating dust grains, or the ongoing replenishment of small, fast-dissipating dust grains.  Under the right conditions, however, dust modeling can break this degeneracy and give us insights into the origin of active asteroid activity (e.g., \textit{Hsieh et al.}, 2004; \textit{Ishiguro et al.}, 2011b; \textit{Agarwal et al.}, 2013).

Pinpointing cases of rotational instability can be difficult because there are few clear predictions about the likely appearances of such bodies.  The repeated dust ejections from 311P, for example, can be reconciled with a rotational shedding instability, but it is quite likely that a range of morphologies and time-dependent emission profiles can be produced by this mechanism.  Significantly, other mechanisms seem incompatible with the observations.  The multiple object P/2013 R3 is self-evidently a result of break-up, with rotation as the likely cause.  

Table \ref{actives} summarizes our conclusions about the mechanisms driving activity in each object.   For many objects, we can eliminate several processes but cannot isolate a unique cause.  This partly reflects the inadequacies of the data but is also a true consequence of the nature of the activity.  For example, in those cases where sublimation is believed to be the primary activity driver, an impact (or other disturbance) may be needed to excavate buried ice to trigger sublimation, and mass loss may be assisted by rapid rotation.  We emphasize that, in many cases, the entries in Table \ref{actives} are expected to change with the acquisition of new data.

\subsection{Distribution} 
The orbital distribution of the active asteroids may provide a clue as to the activity mechanisms.  There is apparent clustering of active asteroids in Figure (\ref{ae}) in the outer belt near $a \sim$ 3.1$\pm$0.1 AU.   Of the 13 active asteroids orbiting in the main-belt, only two (P/2010 A2 and 311P) are located inside the 3:1 mean-motion resonance with Jupiter ($a$ = 2.5 AU), corresponding to $\sim$15\% of the total while, for the asteroids as a whole, this ratio is $\sim$35\%.  However,  the difference between these distributions is not significant. A Kolmogorov-Smirnov test applied to compare the semimajor axes of the active asteroids and the first 500 numbered asteroids shows that the distributions have a 
17\% probability of being drawn from the same parent distribution and, therefore, are consistent.  However, when the comparison is made between the distribution of the semimajor axes of active asteroids and  of  main-belt asteroids of comparable size, the difference instead becomes highly significant.  Specifically, there is  $<$0.1\% likelihood that active asteroids and main-belt asteroids with absolute magnitude $H >$ 18 are drawn from the same parent population, according to the K-S test.  This different result occurs because of observational bias in magnitude-limited sky survey data, which strongly inhibits the detection of small asteroids in the outer belt.  While the active asteroids are subject to this same bias,  their detectability is also influenced by their activity, as is evident from the fact that many objects (e.g.~P/2010 A2, 311P, P/2013 R3) were unknown before being discovered in an active state.  We conclude that the relative paucity of active asteroids in the inner-belt and the concentration near 3.1 AU are real, but larger samples from quantitative sky surveys will be important to better assess the biases.    Clustering could suggest an origin as collisionally produced fragments of an ice-rich precursor (\textit{Novacovik et al.,} 2012, 2014). 

\section{\textbf{DYNAMICS}}
Numerical simulations have been conducted to assess the long-term dynamical stability of many of the known active asteroids (e.g., \textit{Jewitt et al.}, 2009; \textit{Haghighipour}, 2009; \textit{Hsieh et al.}, 2012a,b,c, 2013; \textit{Stevenson et al.}, 2012).  This is of particular interest in the context of understanding whether ice-bearing main-belt comets are native to the asteroid belt or are implanted interlopers from the outer solar system.  In most cases, while the median dynamical lifetime of short-period comets before ejection from the solar system or collision with the Sun was found to be $\sim5\times10^5$~yr (\textit{Levison \& Duncan}, 1994), active asteroids in the asteroid belt have been found to be stable over timescales of 10$^8$~yr or longer.   238P and 259P are notable exceptions, however, having been found to be unstable on timescales on the order of $\sim10^7$~yr (\textit{Jewitt et al.}, 2009; \textit{Haghighipour}, 2009).  

\textit{Fern\'andez et al.} (2002) forward-integrated the orbits of test particles representing the 202 JFCs known at the time and several dynamical clones of each comet (using purely gravitational computations). None ended up on  orbits with both low eccentricity and low inclination like that of 133P.  Comet 503D/Pigott was seen to evolve onto a main-belt orbit, but with high inclination.  Interestingly, we know now of active asteroids in the main belt with high inclinations (e.g., 259P and P/2010 R2), and so the possibility that these could be JFC interlopers cannot be excluded.  In the same work, it was shown that detaching objects from the JFC population was possible with the inclusion of non-gravitational (NG) forces in the simulations, but such large forces were needed as to be unrealistic.  The size dependence of NG forces plays a role in this, since NG forces should be far less effective for larger objects, while smaller objects may not be able to sustain such large NG forces for very long without disintegrating or exhausting their volatile content.  

The aforementioned simulations assume the modern-day architecture of the solar system.  Some dynamical models of the early solar system suggest that planet migration could have resulted in the emplacement of icy outer solar system objects in what is now the main asteroid belt (e.g., \textit{Levison et al.}, 2009; \textit{Walsh et al.}, 2011), although probably not in low inclination orbits like those of many active asteroids.  Therefore, particularly when assessing the astrobiological significance of icy active asteroids as indicators of the composition of the inner protosolar disk, the possibility that these objects could have been implanted at early times must continue to be considered (see the chapter by \textit{Morbidelli et al.} in this volume).

Several active asteroids are associated with collisional asteroid families and clusters. For example, four active asteroids (133P, 176P, 238P, and 288P) have been linked to the Themis family either currently or in the past (\textit{T\'oth}, 2000; \textit{Hsieh}, 2009; \textit{Haghighipour}, 2009; \textit{Hsieh et al.}, 2012c), while P/2010 A2 and 311P are associated with the Flora family (\textit{Moreno et al.}, 2010; \textit{Jewitt et al.}, 2013c) and P/2012 T1 and 313P with the Lixiaohua family (\textit{Hsieh et al.}, 2013, 2015b).  The dynamical associations are consistent with photometry showing, for example, that the colors of 133P, 176P and 238P are close to those of the Themis family (\textit{Licandro et al.,} 2011a) while that of 311P is consistent with membership in the Flora family (\textit{Jewitt et al.} 2013c).  However, dynamical associations are unsurprising, given that many asteroids are family members, and a causal relationship between family membership and activity cannot be assumed.  The color relationships could also be coincidental, since S-types are common in the inner-belt where P/2010 A2 and 311P are found, and C-types in the outer belt where 133P, 176P, 238P, and 288P orbit.  However, several active objects have also been linked to much smaller, young asteroid clusters, including 133P (part of the $<$10~Myr-old Beagle family; \textit{Nesvorn\'y et al.}, 2008), 288P (part of a newly-discovered 7.5$\pm$0.3~Myr-old cluster; \textit{Novakovi\'c et al.}, 2012), and P/2012 F5 (part of a 1.5$\pm$0.1~Myr-old cluster; \textit{Novakovi\'c et al.}, 2014). In the case of active asteroids exhibiting sublimation-driven activity, membership in a young cluster suggests a natural mechanism by which ice could have been preserved over long timescales (i.e., within the interior of a larger parent body) and only recently exposed to more direct solar heating.  In the case of objects exhibiting impact-induced activity, an associated young cluster could represent both a source of potential impactors (given that cluster members will share very similar orbits) and possibly a consequence of residing in a region of the asteroid belt characterized by high collision rates (e.g., \textit{Novakovi\'c et al.}, 2014). 

\section{\textbf{SURVEYS AND RATES}}
About 10 of the known active asteroids (Table \ref{orbital}) and $\sim$10$^6$ main-belt asteroids (\textit{Jedicke and Metcalfe} 1998) are larger than 1 km. The ratio of these numbers gives an active fraction, for all types of activity-driving mechanism, $f \sim $ 10$^{-5}$.  However, most asteroids fall near the limiting magnitudes of the surveys in which they are discovered and so are never effectively searched for dust emission. Therefore, $f \sim $ 10$^{-5}$ sets a strong lower limit to the active fraction, since many active objects must go undetected.  Recent, dedicated surveys have attempted to remedy this situation by searching for comae in well-defined sky surveys.  Unfortunately, most of these surveys are so small in scale that they detect no active objects (of the known active asteroids, only 118401 LINEAR (1999 RE70) was discovered as the result of a targeted survey), allowing 3$\sigma$ upper limits, $f_{3\sigma}$, to be determined.  These include $f_{3\sigma} \le$ 54 per million  (\textit{Gilbert and Wiegert} 2009, 2010), $f_{3\sigma} \le$ 6000 per million (\textit{Sonnett et al.}~2011), and $f_{3\sigma} \le$ 50 per million (\textit{Waszczak et al.}~2014).  These surveys were based on the attempted detection of spatially resolved coma.  \textit{Cikota et al.,} (2014) instead used photometry to search for the brightening produced by the ejection of dust, finding five candidates from a search of $\sim$75 million observations of $\sim$300,000 asteroids.  These candidates remain unconfirmed and it is not clear which method, resolved imaging vs.~integrated photometry, offers the better sensitivity to mass loss.

\textit{Hsieh et al.} (2015a) concluded from a sample of 30,000 objects observed near perihelion (giving two detections) that $f \sim$ 100 per million (in the outer belt), which we take as the best current estimate of the fraction of asteroids which are measurably active at any instant.  This is still a lower limit to the true active fraction, in the sense that the surveys are sensitivity-limited, and it is  likely that more evidence for mass loss could be found by increasing the sensitivity.    \textit{Sonnett et al.~(2011)} hint at this possibility by finding statistical evidence for comae in the averaged profiles of many asteroids, although not in any individual object.  If real, their detection may point to electrostatic or gardening dust losses, or some other process acting broadly across the asteroid belt.  

Of the present sample, at least three active asteroids are likely to be driven by outgassing (133P, 238P, 313P), based on activity repeated in different orbits, corresponding to a fraction $\gtrsim$ 15\%.  The fraction of sublimation-driven objects is therefore of order $f \gtrsim$ 15 per million.  The fraction of asteroids which contain ice, $f_{ice}$, could be much larger.  We write $f_{ice} = f / f_d$, where $f_d$ is the duty cycle, equal to the fraction of time for which the average asteroid loses mass.  In the ice excavation model described earlier, $f_d$ is the ratio of the lifetime of an exposed ice patch to the interval between exposures (caused by boulder impacts?) on the same body.  The present level of ignorance about $f_d$ allows the possibility that $f_{ice}$ = 1, meaning that potentially all asteroids could contain ice without violating the survey limits. This surprising result indicates that the dividing line between comets and asteroids may be considerably less sharp than once supposed. \\

\textbf{ Acknowledgments.} Partial support for this work was provided by NASA through grants from the Space Telescope Science Institute, which is operated by AURA, Inc, and through their Solar System Observations program.

\begin{deluxetable}{llllrcr}
\tablecaption{Summary of Orbital Properties
\label{orbital}}
\tablewidth{0pt}
\tablehead{
\colhead{Name} &\colhead{$T_J$\tablenotemark{a}} & \colhead{$a$\tablenotemark{b}} & \colhead{$e$  \tablenotemark{c}} & \colhead{$i$\tablenotemark{d}} & \colhead{$q$\tablenotemark{e}}   & \colhead{$Q$\tablenotemark{f}} }
\startdata
(3200) Phaethon                & 4.508 & 1.271 & 0.890 & 22.17 & 0.140 & 2.402 \\
311P/PANSTARRS (P/2013 P5)          & 3.662 & 2.189 & 0.115 &  4.97 & 1.936 & 2.441 \\
P/2010 A2 (LINEAR)             & 3.582 & 2.291 & 0.124 &  5.26 & 2.007 & 2.575 \\
(1) Ceres                      & 3.309 & 2.768 & 0.076 & 10.60 & 2.556 & 2.979 \\
(2201) Oljato                  & 3.299 & 2.172 & 0.713 &  2.52 & 0.623 & 3.721 \\
P/2012 F5 (Gibbs)              & 3.228 & 3.004 & 0.042 &  9.73 & 2.877 & 3.129 \\
259P/Garradd (P/2008 R1)    & 3.216 & 2.726 & 0.342 & 15.90 & 1.794 & 3.658 \\
(596) Scheila                  & 3.208 & 2.928 & 0.165 & 14.66 & 2.445 & 3.411 \\
288P/(300163) 2006 VW$_{139}$  & 3.203 & 3.050 & 0.200 &  3.24 & 2.441 & 3.659 \\
(62412) 2000 SY$_{178}$ & 3.197 & 3.146 & 0.090 & 4.76 & 2.864 & 3.445 \\
P/2013 R3 (Catalina-PANSTARRS) & 3.185 & 3.033 & 0.273 &  0.90 & 2.204 & 3.862 \\
133P/(7968) Elst-Pizarro       & 3.184 & 3.157 & 0.165 &  1.39 & 2.636 & 3.678 \\
176P/(118401) LINEAR           & 3.167 & 3.196 & 0.192 &  0.24 & 2.582 & 3.810 \\
238P/Read (P/2005 U1)          & 3.152 & 3.165 & 0.253 &  1.27 & 2.364 & 3.966 \\
P/2012 T1 (PANSTARRS)          & 3.134 & 3.154 & 0.236 & 11.06 & 2.411 & 3.897 \\
313P/Gibbs (P/2014 S4)         & 3.132 & 3.156 & 0.242 & 10.97  & 2.391 & 3.920 \\
P/2010 R2 (La Sagra)           & 3.098 & 3.099 & 0.154 & 21.39 & 2.622 & 3.576 \\
107P/(4015) Wilson-Harrington  & 3.083 & 2.638 & 0.624 &  2.79 & 0.993 & 4.284 \\
\enddata


\tablenotetext{a}{Tisserand parameter with respect to Jupiter} 
\tablenotetext{b}{Semimajor axis [AU]}
\tablenotetext{c}{Orbital eccentricity}
\tablenotetext{d}{Orbital inclination}
\tablenotetext{e}{Perihelion distance [AU]}
\tablenotetext{f}{Aphelion distance [AU]}

\end{deluxetable}

\begin{deluxetable}{llllllcrrr}
\tabletypesize{\small}
\tablecaption{Physical Properties
\label{physical}}
\tablewidth{0pt}
\tablehead{
\colhead{Name} &\colhead{$D$\tablenotemark{a}} & \colhead{$p_V$\tablenotemark{b}} & \colhead{$P$  \tablenotemark{c}} & \colhead{$B-V$  \tablenotemark{d}} & \colhead{$dm/dt$  \tablenotemark{e}}}
\startdata

(3200) Phaethon$^{1}$                 & 5 - 7& 0.08 - 0.17 & 3.603 & 0.58$\pm$0.01 & N/A \\
311P/2013 P5 (PANSTARRS)$^{2}$          & $<$0.5 & 0.29\tablenotemark{f} & ? & 0.77$\pm$0.03 & ? \\
P/2010 A2 (LINEAR)$^{3}$                & 0.12              & 0.1\tablenotemark{f}  & ?               & ?  & N/A       \\
(1) Ceres$^{4}$                      & 975               & 0.090$\pm$0.003       & 9.07            & ? & $\sim$6   \\
(2201) Oljato$^{5}$                   & 1.8               & 0.43$\pm$0.03         & ?               & 0.83$\pm$0.02  &  5? (gas) \\
P/2012 F5 (Gibbs)$^6$                   &  1.8 & 0.05\tablenotemark{f} & 3.24$\pm$0.01 & -- & -- \\
259P/(Garradd)$^{7}$                    & 0.30$\pm$0.02     & 0.05\tablenotemark{f} & ?               & 0.63$\pm$0.03 & $\le$1.5 (gas), 0.01 \\
(596) Scheila$^{8}$                   & 113$\pm$2         & 0.038$\pm$0.004       & 15.848          & 0.71$\pm$0.03 & $\le$3 (gas) \\
288P/(300163) 2006 VW$_{139}$$^{9}$     & 3                 & 0.04\tablenotemark{f} & ?               & ?             & ? \\
(62412) 2000 SY178$^{10}$ & 7.8$\pm$0.6 & 0.065$\pm$0.010 & 3.33$\pm$0.01 & 0.64$\pm$0.03 &  ? \\

P/2013 R3 (Catalina-PANSTARRS)$^{11}$ & $<$0.4 (multiple) & 0.05\tablenotemark{f} & ?               & 0.66$\pm$0.04             & ? \\
133P/(7968) Elst-Pizarro$^{12}$        & 3.8$\pm$0.6       & 0.05$\pm$0.02         & 3.471$\pm$0.001 & 0.65$\pm$0.03 & $<$0.04 (gas), 0.01, 0.7-1.6   \\
176P/(118401) LINEAR$^{13}$            & 4.0$\pm$0.4       & 0.06$\pm$0.02         & 22.23$\pm$0.01  & 0.63$\pm$0.02 & 0.1            \\
238P/Read$^{14}$                         & 0.8               & 0.05\tablenotemark{f} & ?               & 0.63$\pm$0.05 & 0.2            \\
P/2012 T1 (PANSTARRS)$^{15}$                & 2.4                & 0.05\tablenotemark{f}                    & --              & 0.65$\pm$0.07           & --             \\
313P/2014 S4 (Gibbs)$^{16}$ & 1.0 & 0.05\tablenotemark{f} & ? & 0.72$\pm$0.02& 0.2 to 0.4 \\
P/2010 R2 (La Sagra)$^{17}$           & 1.1               & 0.04\tablenotemark{f} & ?               & ?             & 4              \\
107P/(4015)Wilson-Harrington$^{18}$         & 3.5$\pm$0.3       & 0.06$\pm$0.01         & 7.15            & ?             & $\le$150 (gas) \\

\enddata


\tablenotetext{a}{Effective diameter (km)}
\tablenotetext{b}{Geometric albedo}
\tablenotetext{c}{Rotation period}
\tablenotetext{d}{Color index. Solar color is B - V = 0.64$\pm$0.02}
\tablenotetext{e}{Inferred mass loss rate in kg s$^{-1}$. Unless otherwise stated, the estimates are based on continuum measurements and refer to dust.  N/A means that no mass loss rate can be specified because the loss is not in steady state, or for some other reason.}
\tablenotetext{f}{Value is assumed, not measured}
\tablenotetext{1}{Ansdell et al.~2014, $^{2}$Jewitt et al.~2013c, 2014c, Hainaut et al.~2014, $^{3}$Jewitt et al.~2010, 2011a$^{4}$Kuppers et al., 2014, $^{5}$Tedesco et al.~2004, McFadden et al.~1993, Russell et al.~1984, $^{6}$Stevenson et al.~2012, Moreno et al.~2012, Novakovic et al.~2014, Drahus et al.~2015 $^{7}$Jewitt et al.~2009, $^{8}$Tedesco et al.~2002,  Warner 2006, $^{9}$Hsieh et al.~2011d, $^{10}$Sheppard and Trujillo 2014,  $^{11}$Jewitt et al.~2014b $^{12}$Hsieh et al.~2004, 2009a, 2011a, $^{13}$Hsieh et al.~2011, Licandro et al.~2011, $^{14}$Hsieh et al.~2011c, $^{15}$ Moreno et al.~2013, Hsieh et al.~2013, O'Rourke et al.~2013 $^{16}$Jewitt et al. 2015, Hui and Jewitt 2015, $^{17}$Moreno et al.~2011b, Hsieh et al.~2012c, Hsieh 2014 $^{18}$Veeder et al.~1984, Fernandez et al.~1997,  Licandro et al.~2009, Urakawa et al.~2011, Ishiguro et al.~2011b. 
}

\end{deluxetable}

\begin{deluxetable}{lccccc}
\tabletypesize{\small}
\tablecaption{Summary of Mechanisms\tablenotemark{a}
\label{actives}}
\tablewidth{0pt}
\tablehead{
\colhead{Name} &\colhead{Sublimation} & \colhead{Impact} & \colhead{Electrostatics} & \colhead{Rotation} & \colhead{Thermal} }
\startdata

(3200) Phaethon                & $\times$     & ?            & ?        & ?            & $\checkmark$ \\
311P/PANSTARRS (P/2013 P5)     & $\times$     & $\times$     & $\times$ & $\checkmark$ & $\times$     \\
P/2010 A2 (LINEAR)             & $\times$     & $\checkmark$ & $\times$ & $\checkmark$ & $\times$     \\
(1) Ceres                      & $\checkmark$ & $\times$     & $\times$ & $\times$     & $\times$     \\
(2201) Oljato                  & ?            & ?            & ?        & ?            & $\times$     \\
P/2012 F5 (Gibbs)              & $\times$     & $\checkmark$ & $\times$ & $\checkmark$             & $\times$     \\
259P/Garradd (P/2008 R1)       & ?            & ?            & ?        & ?            & $\times$     \\
(596) Scheila                  & $\times$     & $\checkmark$ & $\times$ & $\times$     & $\times$     \\
288P/(300163) 2006 VW$_{139}$  & $\checkmark$ & ?            & ?                 & ?  & $\times$    \\
(62412) 2000 SY$_{178}$ & ? & $\checkmark$ & ? & $\checkmark$ &  $\times$ \\

P/2013 R3 (Catalina-PANSTARRS) & ? & $\times$     & $\times$ & $\checkmark$ & $\times$     \\
133P/(7968) Elst-Pizarro       & $\checkmark$ & $\times$     & ?        & $\checkmark$            & $\times$     \\
176P/(118401) LINEAR           & $\checkmark$ & ?            & ?        & $\times$     & $\times$     \\
238P/Read (P/2005 U1)             & $\checkmark$ & $\times$     & $\times$ & ?            & $\times$     \\
P/2012 T1 (PANSTARRS)          & $\checkmark$ & $\times$     & $\times$ & ?            & $\times$     \\
313P/Gibbs (P/2014 S4)            &  $\checkmark$ & $\times$ & $\times$ & ? & $\times$ \\
P/2010 R2 (La Sagra)           & $\checkmark$ & ?            & ?        & ?            & $\times$     \\
107P/(4015) Wilson-Harrington  & ?            & ?            & ?        & $\times$     & $\times$     \\

\enddata

\tablenotetext{a}{$\checkmark$ - evidence exists consistent with the process, $\times$ - evidence exists inconsistent with the process, $?$ - insufficient or mixed evidence exists }

\end{deluxetable}

\end{document}